\documentclass[pra,10pt,aps,superscriptaddress,twocolumn,footinbib,notitlepage,tightenlines]{revtex4-2}
\usepackage{mathtools}
\usepackage{amsmath}
\usepackage[shortlabels]{enumitem}
\usepackage{graphicx,epic,times,eepic,epsfig,latexsym,verbatim,color}

\usepackage{amsfonts}       
\usepackage{nicefrac}       
\usepackage[linesnumbered,ruled,procnumbered]{algorithm2e}
\usepackage[table]{xcolor}

\usepackage{framed}
\usepackage{bbm}
\usepackage{caption}

\usepackage{float}
\usepackage{tikz}
\usetikzlibrary{chains}
\usetikzlibrary{fit}
\usepackage{epsfig}
\usetikzlibrary{shapes.symbols,patterns} 
\usepackage{pgfplots}
\usepackage[strict]{changepage}
\usepackage[marginal]{footmisc}
\usepackage{url}
\usepackage{theorem}

\newtheorem{proposition}{Proposition}
\newtheorem{lemma}[proposition]{Lemma}

\newtheorem{theorem}[proposition]{Theorem}
\newtheorem{corollary}[proposition]{Corollary}

\def\squareforqed{\hbox{\rlap{$\sqcap$}$\sqcup$}}
\def\qed{\ifmmode\squareforqed\else{\unskip\nobreak\hfil
\penalty50\hskip1em\null\nobreak\hfil\squareforqed
\parfillskip=0pt\finalhyphendemerits=0\endgraf}\fi}
\def\endenv{\ifmmode\;\else{\unskip\nobreak\hfil
\penalty50\hskip1em\null\nobreak\hfil\;
\parfillskip=0pt\finalhyphendemerits=0\endgraf}\fi}
\newenvironment{proof}{\noindent \textbf{{Proof~} }}{\hfill $\blacksquare$}
\newcounter{remark}
\newenvironment{remark}[1][]{\refstepcounter{remark}\par\medskip\noindent%
\textbf{Remark~\theremark #1} }{\medskip}
\newcounter{example}

\mathchardef\ordinarycolon\mathcode`\:
\mathcode`\:=\string"8000
\def\vcentcolon{\mathrel{\mathop\ordinarycolon}}
\begingroup \catcode`\:=\active
  \lowercase{\endgroup
  \let :\vcentcolon
  }

\definecolor{darkblue}{RGB}{0,76,156}
\definecolor{darkkblue}{RGB}{0,0,153}
\definecolor{blue2}{RGB}{102,178,255}
\definecolor{darkred}{RGB}{195,0,0}

\newcommand{\nc}{\newcommand}
\nc{\rnc}{\renewcommand}
\nc{\lbar}[1]{\overline{#1}}
\nc{\bra}[1]{\langle#1|}
\nc{\ket}[1]{|#1\rangle}
\nc{\dketbra}[2]{\vert #1 \rangle \hspace{-.8mm} \rangle \hspace{-.4mm} \langle\hspace{-.8mm}\langle #2 \vert}
\nc{\dbra}[1]{\langle\hspace{-.8mm}\langle #1\vert}
\nc{\dket}[1]{\vert#1\rangle\hspace{-.8mm}\rangle}
\nc{\ketbra}[2]{|#1\rangle\!\langle#2|}
\nc{\braket}[2]{\langle#1|#2\rangle}
\newcommand{\braandket}[3]{\langle #1|#2|#3\rangle}

\nc{\proj}[1]{| #1\rangle\!\langle #1 |}
\nc{\avg}[1]{\langle#1\rangle}
\nc{\rank}{\operatorname{Rank}}
\nc{\smfrac}[2]{\mbox{$\frac{#1}{#2}$}}
\nc{\tr}{\operatorname{Tr}}
\nc{\ox}{\otimes}
\nc{\dg}{\dagger}
\nc{\dn}{\downarrow}
\nc{\cA}{{\cal A}}
\nc{\cB}{{\cal B}}
\nc{\cC}{{\cal C}}
\nc{\cD}{{\cal D}}
\nc{\cE}{{\cal E}}
\nc{\cF}{{\cal F}}
\nc{\cG}{{\cal G}}
\nc{\cH}{{\cal H}}
\nc{\cI}{{\cal I}}
\nc{\cJ}{{\cal J}}
\nc{\cK}{{\cal K}}
\nc{\cL}{{\cal L}}
\nc{\cM}{{\cal M}}
\nc{\cN}{{\cal N}}
\nc{\cO}{{\cal O}}
\nc{\cP}{{\cal P}}
\nc{\cQ}{{\cal Q}}
\nc{\cR}{{\cal R}}
\nc{\cS}{{\cal S}}
\nc{\cT}{{\cal T}}
\nc{\cU}{{\cal U}}
\nc{\cV}{{\cal V}}
\nc{\cX}{{\cal X}}
\nc{\cY}{{\cal Y}}
\nc{\cZ}{{\cal Z}}
\nc{\cW}{{\cal W}}
\nc{\csupp}{{\operatorname{csupp}}}
\nc{\qsupp}{{\operatorname{qsupp}}}
\nc{\var}{{\operatorname{var}}}
\nc{\rar}{\rightarrow}
\nc{\lrar}{\longrightarrow}
\nc{\polylog}{{\operatorname{polylog}}}
\nc{\wt}{{\operatorname{wt}}}
\nc{\av}[1]{{\left\langle {#1} \right\rangle}}
\nc{\supp}{{\operatorname{supp}}}
\nc{\VComb}{{\widetilde{\cal C}}}
\nc{\VChoi}{{\widetilde{C}}}
\nc{\idop}{{\mathbbm{1}}}
\nc{\argmin}{{\operatorname{argmin}}}

\def\a{\alpha}

\nc{\RR}{{{\mathbb R}}}
\nc{\CC}{{{\mathbb C}}}
\nc{\FF}{{{\mathbb F}}}
\nc{\NN}{{{\mathbb N}}}
\nc{\ZZ}{{{\mathbb Z}}}
\nc{\PP}{{{\mathbb P}}}
\nc{\QQ}{{{\mathbb Q}}}
\nc{\UU}{{{\mathbb U}}}
\nc{\EE}{{{\mathbb E}}}
\nc{\id}{{\operatorname{id}}}

\nc{\CHSH}{{\operatorname{CHSH}}}

\nc{\abs}[1]{\left\lvert {#1} \right\rvert}
\nc{\norm}[1]{\left\lVert {#1} \right\rVert}
\nc{\ceil}[1]{\left\lceil {#1} \right\rceil}

\usepackage{hyperref}
\hypersetup{colorlinks=true,citecolor=blue,linkcolor=blue,filecolor=blue,urlcolor=blue,breaklinks=true}
\usepackage{cleveref}

\usepackage{thmtools}
\usepackage{thm-restate}
\usepackage{etoolbox}

\usepackage{blkarray}
\usepackage{amssymb,array,multirow,bm,tcolorbox,relsize,booktabs}
\usepackage[utf8]{inputenc}
\usepackage[T1]{fontenc}
\usepackage{extarrows}
\usepackage{qcircuit}

\definecolor{colortwo}{rgb}{0.4,0.77,0.17}
\definecolor{colorthree}{rgb}{0.01,0.51,0.93}

\pgfplotsset{compat=1.18} 

\usepackage{soul}
\soulregister\cite7
\soulregister\ref7
\soulregister\pageref7
\nc{\LZ}[1]{\textcolor{colortwo}{(Lei: #1)}}
\nc{\XW}[1]{\textcolor{magenta}{(XW: #1)}}
\nc{\YJ}[1]{\textcolor{colorthree}{(YJ: #1)}}
\nc{\YA}[1]{\textcolor{cyan}{(YA: #1)}}

\nc{\add}[1]{{\color{red} #1}}
\nc{\remove}[1]{{\color{lightgray} \st{#1}}}

\renewcommand{\add}[1]{#1}
\renewcommand{\remove}[1]{}

\nc{\replace}[2]{\remove{#1} \add{#2}}

\nc{\Uin}{ U }
\nc{\su}{ \operatorname{SU} }
\nc{\aux}{ \text{aux} }
\nc{\bmP}{ \mathbf{P} }
\nc{\bmF}{ \mathbf{F} }
\nc{\bmV}{ \mathbf{V} }
\nc{\ft}{ \operatorname{FT} }
\nc{\SEL}{{\operatorname{SEL}}}

\nc{\trace}[2][]{ \tr_{#1}\left[ #2 \right] }
\nc{\integral}[2][]{ \int_{#1} \text{d}{#2}\, }

\nc{\amp}[1]{ \cA_{#1} }
\nc{\encod}{ E }
\nc{\decod}{ D }

\nc{\Span}[1]{\operatorname{Span} \left( #1 \right)}

\nc{\set}[1]{ \left\{ #1 \right\} }
\nc{\setcond}[2]{ \left\{ #1 : #2 \right\} }
\begin{document}
\title{Quantum Algorithm for Reversing Unknown Unitary Evolutions}

\author{Yu-Ao Chen}
\author{Yin Mo}
\author{Yingjian Liu}
\author{Lei Zhang}
\author{Xin Wang}
\email{felixxinwang@hkust-gz.edu.cn}
\affiliation{Thrust of Artificial Intelligence, Information Hub,\\
The Hong Kong University of Science and Technology (Guangzhou), Guangzhou 511453, China}

\begin{abstract}
Reversing an unknown quantum evolution is of central importance to quantum information processing and fundamental physics, yet it remains a formidable challenge as conventional methods necessitate an infinite number of queries to fully characterize the quantum process. Here we introduce the Quantum Unitary Reversal Algorithm (QURA), a deterministic and exact approach to universally reverse arbitrary unknown unitary transformations using $\mathcal{O}(d^2)$ calls of the unitary, where $d$ is the system dimension. Our quantum algorithm resolves a fundamental problem of time-reversal simulations for closed quantum systems by confirming the feasibility of reversing any unitary evolution without knowing the exact process. The algorithm also provides the construction of a key oracle for unitary inversion in many quantum algorithm frameworks, such as quantum singular value transformation. It notably reveals a sharp boundary between the quantum and classical computing realms and unveils a quadratic quantum advantage in computational complexity for this foundational task.
\end{abstract}

\date{\today}
\maketitle


\begin{figure*}[t]
\centering
\includegraphics[width=0.85\linewidth]{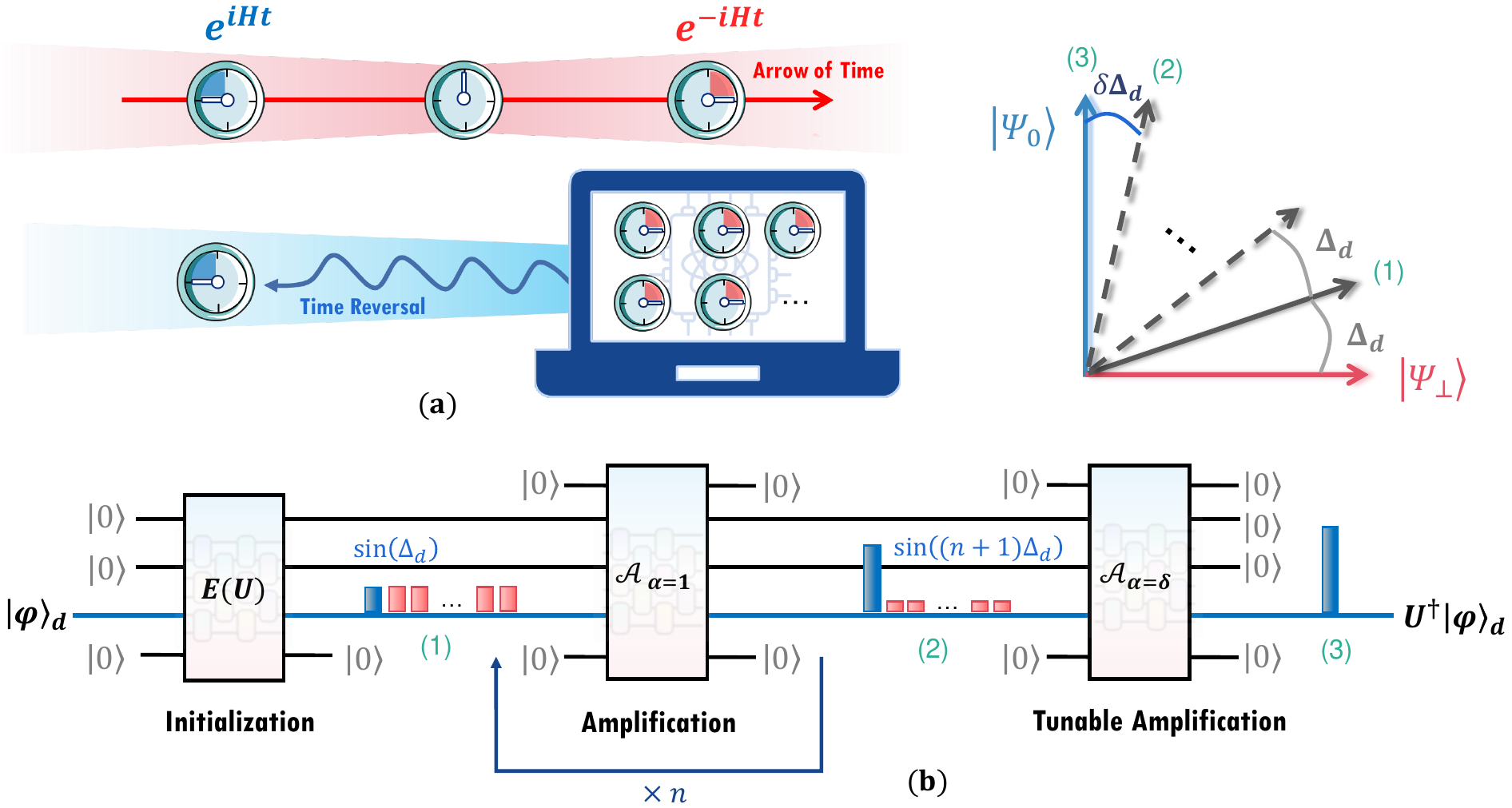}
\caption{Schematic depiction of Quantum Unitary Reversal Algorithm.
(a). A quantum computer could simulate the inverse of an unknown unitary evolution by querying it for finite times.
(b).  QURA, akin to the idea of amplitude amplification, is structured in three stages (1), (2), and (3). The input quantum state $\ket{\varphi}_d$ is initialized into the superposition of the target $U^{\dag} \ket{\varphi}_d$ and the other unwanted components with averaged amplitudes. 
A duality-based amplitude amplifier $\amp{\alpha=1}$ is used iteratively to enhance the amplitude of the target state with a constant angle $\Delta_d$. The iteration ends when the angle approaches its maximum before surpassing $\pi/2$. Consequently, a tunable amplifier, denoted as $\amp{\alpha=\delta}$, completes the angular amplification up to $\pi/2$, giving rise to the inverse of unitary with all ancillary qudits returning to zero states. }
\label{fig:unitary inv}
\end{figure*}

\emph{Introduction.---} 
Can we reverse the evolution of a quantum system without any knowledge of its specifics?
{While it is easy to reverse a unitary with known construction, how to reverse an unknown unitary has remained a long-standing open problem. If possible, it plays a pivotal role within quantum computing and would offer a generalization that extends the capability and practicality of many quantum algorithms~\cite{wang2023quantum,odake2023universal}. Among these, the quantum singular value transformation (QSVT)~\cite{gilyen2019quantum} emerges as a particularly prominent and comprehensive framework, which requires the implementation of a block-encoded unitary as well as its inverse for polynomial manipulation of the encoded data. Here we give an affirmative response to this intriguing question by presenting} 
the first general protocol that could reverse the arbitrary evolution of closed quantum systems by querying the evolution for finite times.
This advancement not only opens up new avenues for quantum algorithm design but also sheds light on deeper questions in fundamental physics, such as the nature of time reversal in quantum systems and the interplay between quantum mechanics and information theory.

{While conventional classical approaches, such as quantum process tomography~\cite{Baldwin2014, Gutoski2014, Mohseni2008, Haah2023}, to reversing unknown unitaries encounter significant limitations in terms of efficiency and scalability in characterizing the unknown process, the quantum computer offers a new possibility by leveraging the original forward evolution process itself and shedding light on demonstrating quantum advantages via quantum simulation~\cite{Feynman1982}. The idea of performing targeted quantum tasks without needing classical information is fascinating and potentially transformative, where quantum teleportation protocol~\cite{bennett1992communication, bennett1993teleporting} is a well-known example that} not only plays a crucial role in quantum communication but also has been pivotal in the development of fault-tolerant quantum computing~\cite{shor1996fault}.

{For the unitary inversion task, }
Ref.~\cite{Quintino2018, Quintino2019, yang2021representation, trillo2023universal} provide positive evidence in this direction by showing a universal probabilistic heralded quantum circuit that implements the exact inverse of an unknown unitary. Notably, Ref.~\cite{Yoshida2023} took a further step by proposing \replace{a}{the first} deterministic and exact protocol to reverse any unknown qubit unitary transformation and {recently Ref.~\cite{mo2025parameterized} discovers a simplier protocol via parameterized circuit architecture.}
In higher dimensions beyond simple qubit cases, Ref.~\cite{gavorova2024topological, Odake2024} studied the lower bounds on the number of queries to complete the task. 
{While previous results were derived mainly based on SDP optimization, whether qubit unitary is the only special case that can be reversed deterministically and exactly remains a significant open question in quantum information science.}

In this article, we introduce the first universal protocol to reverse any unknown unitary evolution of arbitrary dimension deterministically and exactly with finite calls of the unitary. The protocol, named the Quantum Unitary Reversal Algorithm (QURA), realizes the time reversal of an arbitrary unknown $d$-dimensional closed quantum system by querying its time-evolution operators a mere $\cO(d^2)$ times. This result establishes that a finite number of uses of an unknown unitary suffices to realize its inverse exactly and deterministically, standing in stark contrast to the infinite number of applications required by conventional methods. Notably, the query complexity of QURA is proven to be optimal, as evidenced by the recently developed lower bound~\cite{Odake2024}. 
In constructing this algorithm, a novel amplitude amplification approach is developed, which does not require the oracle to perform the inversion of the target unitary.
Beyond resolving the fundamental question of reversibility of unknown unitary operations, we further show that reversing an unknown unitary on a quantum computer yields a quadratic quantum advantage in computational complexity compared to classical methods that rely on process tomography, and the dependency of process error in query complexity is removed. With this protocol, our work provides the construction of a crucial oracle for unitary inversion within quantum algorithm frameworks, representing a significant step in the quest to harness the full potential of quantum computing.

\emph{Quantum Unitary Reversal Algorithm.---}To show that QURA can achieve the inverse of a unitary operation without knowledge of its specifics, we are going to show the process and the underlying ideas of QURA. 
{The main idea is to first encode and transform the information of the gate $U$ into a dedicated subspace that corresponds to its inversion $U^{-1}$, which will then be amplified to amplitude $1$ to achieve the exact time reversal. 
The amplitude amplification here is essentially different from conventional approaches~\cite{Brassard2002}, which is constructed based on the relationship that the reverse of $U \in \su(d)$ is also its transpose conjugate as $U^{-1} = U^{\dag}$. We name it duality-based amplitude amplification and will show its detail after introducing the whole process of QURA. 
Overall, the inverse of $U$ can be achieved through such a determined quantum circuit is formulated as the following theorem:}

\begin{theorem}\label{thm:unitary inv cir exist}
For any dimension $d$, there exists a quantum algorithm that exactly and deterministically implements $U^\dag$ for arbitrary $d$-dimensional unitary operator $U$, by using at most $d \ceil{\pi / 2 \Delta_d} - 1$ calls of $U$, where $\Delta_d = \arcsin{1/d}$.
\end{theorem}

An intuitive streamline of QURA is depicted in Figure~\ref{fig:unitary inv}, where the key element is shown by the vertical blue bar, corresponding to the amplitude of the target quantum state in the circuit, which we denote as $\ket{\Psi_0}$: 
\begin{equation}\label{eqn:Psi0}
    \ket{\Psi_0} = \ket{0}_{\textrm{anc}} \ox U^\dag\ket{\varphi}
,\end{equation}
where {$\ket{0}_{\textrm{anc}}$ denotes the zero state of ancilla systems}.
The amplitude of state $\ket{\Psi_0}$ will gradually tend to 1 as the algorithm progresses, at which point $U^{\dag}$ is deterministically realized by tracing out the ancillary qudits.

The algorithm is
{realized by first initializing the state to be} the target state $\ket{\Psi_0}$ with some unwanted orthogonal component $\ket{\Psi_{\perp}}$, respective to the vector (1) in Figure~\ref{fig:unitary inv}:
\begin{equation}\label{eqn:Encoder}
\encod(U) \left(\ket{0}_{\textrm{anc}} \ox \ket{\varphi}\right)
= \sin{\left(\Delta_d\right)} \ket{\Psi_0} + \cos{\left(\Delta_d\right)} \ket{\Psi_{\perp}}
\end{equation}
The initialization is realized based on a subcircuit we name as ``intrinsic encoder'' $\encod(U)$, which queries gate $U$ to transform the input state to a dedicated subspace corresponding to $U^{\dag}$. The properties of $\encod(U)$ and how to construct it are shown in supplementary material.

{After initialization,} QURA amplifies the ``good'' state $\ket{\Psi_0}$ 
{using the} ``duality-based amplitude amplifier'' $\amp{\a}$. For an input state $\ket{\Psi_{\theta}}= \sin{\theta} \ket{\Psi_0} + \cos{\theta}\ket{\Psi_{\perp}}$, $\amp{\a}$ will rotate the angle $\theta$ of the state into 
\begin{equation}\label{eqn:amplify}
    \theta \xrightarrow{\amp{\alpha}} \arcsin(\alpha \sin{\theta})+\Delta_d
,\end{equation}
where arbitrary angle from $\Delta_d$ to $\theta+\Delta_d$ could be amplified by selecting an appropriate $\alpha \in \left[0, 1\right]$. It is noticed that for the special case when $\alpha = 1$, $\amp{\a}$ turns to rotate the angle $\theta$ with a constant $\Delta_d$. Hence, amplifier $\amp{\alpha=1}$ is executed for $n = \lceil \pi/{2\Delta_d} \rceil - 2$ times after the initialization to amplify the amplitude from $\sin{\Delta_d}$ into $\sin{((n+1)\Delta_d)}$.
Subsequently, an additional tunable amplifier $\amp{\alpha=\delta}$ with $\delta=\cos\left(\Delta_d\right) / \sin\left((n+1)\Delta_d\right)$ is required to compensate the residue into $\pi/2$.
Since $\amp{\alpha=1}$ works as a constant amplifier, independent of the former angle $\theta$, the tunable amplifier is flexible to move into any section after the initialization. 

The detailed procedure for QURA is shown in Algorithm~\ref{alg:qura}.\remove{, where the specific circuit implementations for each subcircuit can be found in the supplementary material} 
\add{In this algorithm, Step 1 introduces the subcircuits needed for the algorithm, and the implementations of which can be found in the supplementary material~\footnote{See supplementary material for more details.};
Step 2 prepares the state that encode the good state; 
Step 3 - 4 amplifies the amplitude of the good state to 1 via the amplification opertator $\amp{\a=1}$.}

\begin{algorithm}[htbp]
\SetKwInOut{Input}{Inputs}
\SetKwInOut{Output}{Output}
\Input{A quantum state $\ket{\varphi}$, a forward evolution operator $U$.} 
\Output{The state after backward evolution, $\ket{\Psi_0} = U^\dag \ket{\varphi}$.} 

Query $U$ to prepare subcircuits $\encod(U), \amp{\a=1}, \amp{\a=\delta}$, where $\delta = \cos \left(\Delta_d\right)/\sin\left((n+1)\Delta_d\right)$\;

Apply $\encod(U)$ to the input state $\ket{0}_\textrm{anc} \ox \ket{\psi}$ gives $\ket{\psi^{(0)}} \gets \sin{\left(\Delta_d\right)} \ket{\Psi_0} + \cos{\left(\Delta_d\right)} \ket{\Psi_{\perp}}$\;

\For{$n$ from $1$ to $N = \ceil{{\pi}/{ {2\Delta_d}}} - 2$}{
    Apply $\amp{\a=1}$ to $\ket{\psi^{(n - 1)}}$ gives $\ket{\psi^{(n)}} \gets \sin{\left((n+1)\Delta_{d}\right)} \ket{\Psi_0} + \cos{\left((n+1)\Delta_{d}\right)} \ket{\Psi_{\perp}}$\;
}

Output $\amp{\a=\delta} \ket{\psi^{(N)}} = \ket{\Psi_0}$\;

\caption{Quantum Unitary Reversal Algorithm}
\label{alg:qura}
\end{algorithm}

QURA can be understood as a hybrid digital-analog quantum algorithm that simulates the unitary inverse via plugging the original unitary evolution into quantum circuits with open slots. 
\add{We also note that Algorithm~\ref{alg:qura} generalizes the single-qubit unitary inversion protocol purposed in~\cite{mo2025parameterized}, which solves the long-standing open problem of unitary inversion problem for arbitrary dimension.}
This formalism validates the strong expressivity of the quantum comb framework~\cite{chiribella2008quantum} in exploring the quantum functionalities of quantum processes. In particular, QURA serves as a critical building block for realizing the full potential of quantum computing, providing ground support for the practicality of many quantum algorithms, including Grover's algorithm~\cite{Grover1996} and the algorithmic framework of QSVT~\cite{gilyen2019quantum}.

\emph{Duality-based amplitude amplification.---} {From the detailed procedure of QURA, one can see that the core idea is to transform the input state to the 2-dimension subspace, corresponding to Equation~\eqref{eqn:Encoder}, and subsequently amplify the target component to finally realize $U^{\dag}$. When $\a = 1$, \add{$\amp{\a}$ can be represented as
\begin{equation}
    \amp{\a} \equiv \begin{blockarray}{ccc}
      \ket{\Psi_\perp} & \ket{\Psi_0} & \\
      \begin{block}{[cc]c}
        \cos(\Delta_d) & -\sin(\Delta_d) &  \bra{\Psi_\perp}\\
        \sin(\Delta_d) & \cos(\Delta_d) & \bra{\Psi_0} \\
      \end{block}
    \end{blockarray}
.\end{equation}
One may relate such representation with the oblivous amplitude amplification~\cite{berry2014exponential}.}
However, we point out that the amplitude amplification in QURA is \remove{essentially} different from conventional approaches, which require the inversion of the target unitary. \remove{Differently,} $\amp{\a}$ is realized by constructing two modules, ``intrinsic encoder'' $\encod(U)$ and ``shifted decoder'' $\decod(U)$, based on mathematical relationships totally. Here we briefly introduce the idea behind constructing these two modules, which we believe will provide new ideas for the design of quantum algorithms. }

The first thing one could notice is that realizing $U^{\dag}$ could be decomposed into realizing complex conjugate and transpose separately. Notice that the complex conjugate of an arbitrary unitary matrix $U \in \su(d)$ can be achieved using $d-1$ instances of this gate in parallel~\cite{Miyazaki2019}, the unitary inverse could then be realized through the following relationship
\add{
\begin{equation}\label{eqn:transpose_main}
     U^\dag = (U^*)^T= \frac{1}{d} \sum_{j, k = 0}^{d - 1} Z^j X^k U^* Z^{-j} X^k  ,
\end{equation}
}
where {$Z$ and $X$} are the clock and shift operators, respectively. 
{Based on Equation~\eqref{eqn:transpose_main}, we use the idea of linear combination of unitaries~\cite{childs2012hamiltonian}  to construct $\encod(U)$ through querying gate $U$. It enables us to encode the input state to the 2-dimension subspace containing $U^{\dag}$. }

\begin{table*}
\centering
\caption{The table of quantum and classical protocols to realize unitary inversion. Here the error is measured in trace norm distance, and $d$ is the dimension of the input unitary. The second row corresponds to the probabilistic quantum method~\cite{Quintino2018}, where $k$ is the number of queries of $U$ in it; the required resources for the classical method are based on the studies in~\cite{Baldwin2014,Gutoski2014,Mohseni2008,Haah2023, Haah2017b,ODonnell2017,Williams2024}.}~\label{tab:advantage}
\setlength{\tabcolsep}{1em}
\resizebox{\linewidth}{!}{
\begin{tabular}{@{} lccccc @{}}
\toprule
Method & Error & Success Probability & Computational complexity & Query & Number of ancillas\\
\midrule
\addlinespace
QURA (\textbf{this work}) & $0$ & $1$ & $\cO\left(d\right)$ & $\cO\left(d^2\right)$ & $\cO\left(d \log(d)\right)$ qubits \\
\addlinespace
    Ref.~\cite{Quintino2018} 
    & $0$ & $1-(1-1/d^2)^{\lfloor(k+1)/d\rfloor}$
    & $\cO \left( k/d \right)$
    & $k$
    & $\cO\left(d \log(d)\right)$ qubits \\
\addlinespace
    Ref.~\cite{Quintino2018} (take $k = \cO(d^3)$) 
    & $0$ & $\cO(1)$
    & $\cO \left( d^2 \right)$
    & $\cO \left( d^3 \right)$
    & $\cO\left(d \log(d)\right)$ qubits \\
\addlinespace
    Tomo $U$ and prepare $U^\dag$ 
    & $\epsilon$ & $1$
    & $\cO \left(d^2/\epsilon \right)$
    & $\cO \left(d^2/\epsilon \right)$
    & $\Theta(d^2)$ cbits \\
\addlinespace
    Tomo $U$ and $\rho$, and prepare $U^\dag\rho U$ 
    & $\epsilon$ & $1$
    & $\cO\left(d^2/\epsilon^2 + d^{2.37}\right)$
    & $\cO\left(d^2/\epsilon^2\right)$
    & $\Theta\left(d^2\right)$ cbits \\
\bottomrule
\end{tabular}
}
\end{table*}

\begin{figure}[H]
\centering
\includegraphics[width=1\linewidth]{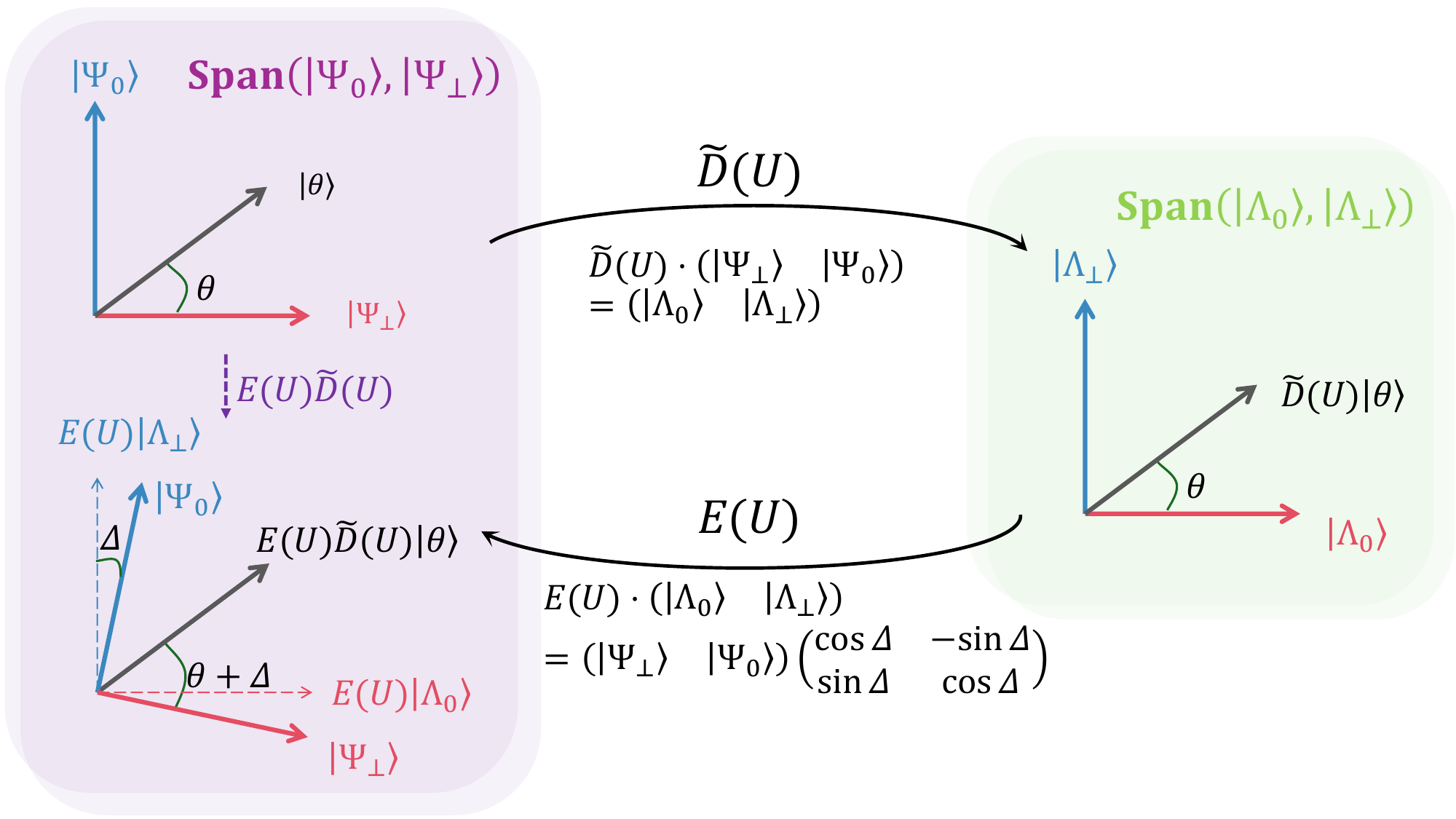}
\caption{\add{An abstract depiction of one rotation of angle $\Delta_d$, between two vector subspaces in $\CC^{2d^3}$ determinated by vectors $\ket{\Psi_0}$, $\ket{\Lambda_0}$ and unitary $\encod(U)$. Here $\widetilde\decod(U)$ is a unitary analog to $\encod(U)^\dag$, constructed by $\decod(U)$.}}
\label{fig:duality}
\end{figure}

To amplify the target component in this 2-dimension subspace, we construct another module $\decod(U)$, which satisfies the following dual relationship with $\encod(U)$:
\begin{align}
\encod(U)\cdot(\ket{\,0\,0\,}\ox\idop)&=\decod(U)^\dag\cdot(\ket{{+}{+}}\ox\idop),\\
\decod(U)\cdot(\ket{{+}{+}}\ox\idop)&=\encod(U)^\dag\cdot(\ket{\,0\,0\,}\ox\idop)
,\end{align}
where $\ket{0}$ and $\ket{+}$ are $d$-dimensional states. This relationship ensures that after running the circuit with this pair of modules, the vectors in this 2-dimensional subspace will remain in it and be rotated with a fixed angle\add{, as summarized in Figure~\ref{fig:duality}}. This is the reason we name the whole subcircuit as the ``duality-based amplitude amplifier''.
By simply adjusting a parameter\remove{ in the amplifier $\amp{\a}$}, the tunable amplification can be realized as the desired angle of rotation can be achieved. The details of the mathematical derivation and how to construct these modules are deferred to the supplementary material. 

At this point, we have introduced the Quantum Unitary Reversal Algorithm (QURA), an approach that harnesses the essence of quantum dynamics and employs a novel duality-based amplitude amplifier to reverse any unknown unitary transformation with a finite number of calls to the original forward evolution. {This seminal result resolves a fundamental question in quantum physics, demonstrating that the precise reversal of unknown unitary evolutions can be realized without knowing its specifics.}

\emph{Advantage over classical methods.--}
Having demonstrated that finite resources suffice to reverse any unknown unitary evolution, a natural question arises: do quantum computers hold a quantum advantage in this task? To address this, We present a complexity analysis comparing QURA with conventional classical approaches for reversing unknown evolutions. The quantity we focus on is the computational complexity, i.e., the overall computational time of the algorithm. We are going to show a {polynomial} quantum advantage over classical methods.

Conventional approaches require obtaining classical information (e.g., the matrix form) through quantum process tomography and storing this information on a classical computer. The computational complexity of {performing tomography on} such a unitary with an $\epsilon$-error in the diamond norm is $\cO(d^2/\epsilon)$~\cite{Baldwin2014, Gutoski2014, Mohseni2008, Haah2023}.  Then implementation of reversed evolution would require $\Theta(d^2)$ experiments to retrieve this process classically, and $\cO(d^2)$ time cost to prepare the operation. If the goal is to obtain the whole matrix of $U \rho U^\dag$, one must also perform the tomography on the state $\rho$, which requires $\cO(d^2/\epsilon^2)$ experiments to achieve $\epsilon$ accuracy~\cite{Haah2017b, ODonnell2017}. This is followed by a $\cO(d^{2.37})$ time cost to compute state $U \rho U^\dag$ using the best matrix multiplication algorithm \cite{Williams2024}.

In contrast, QURA only requires running the ``duality-based amplitude amplifier'' $\cO(d)$ times, with $d-1$ instances of evolution operator being employed in parallel to realize the intrinsic encoder $E(U)$. Then the computational complexity in QURA is $\cO(d)$ and hence achieves a quadratic computational speedup towards its classical counterparts. 
Further, the total query complexity in QURA is $\cO(d^2)$. This complexity is optimal as demonstrated by the lower bound in~\cite{Odake2024}. This result also highlights a significant contrast with {conventional} approaches as {extractions of classical information} cannot achieve zero-error implementation due to their query complexity being inversely proportional to $\epsilon$. These two distinctions underscore the quantum advantage in processing quantum data using quantum computing over classical computing. 

The results of these comparisons are summarized in Table~\ref{tab:advantage}, where
we also provide comparisons with the existing quantum algorithms for reversing unitaries~\cite{Quintino2018}. 
For large $d$, to better compare with our scheme, we find that to ensure the successful {probability} in $\cO(1)$, the number of queries is at least in $\cO(d^3)$, and the computational complexity will be in $\cO(d^2)$. 
Details are deferred to the supplementary material. 
These comparisons suggest that QURA {achieves a} quantum advantage in the forthcoming era of early fault-tolerant quantum computing.

\emph{Concluding remarks.--} We have resolved the open problem of deterministically and exactly reversing unknown unitary evolutions, by introducing the Quantum Unitary Reversal Algorithm, which works for arbitrary dimensional quantum systems. This {development} enhances our capability to execute complex quantum operations without prior knowledge of the underlying processes, marking a {notable step forward} in the field of quantum computing.
By harnessing the inherent dynamics of quantum systems and developing novel duality-based amplitude amplification techniques, we have shown that QURA can universally reverse any unitary transformation with perfect fidelity and success probability, using an optimal $\cO(d^2)$ calls to the unitary and an execution time of $\cO(d)$. 
A key technical contribution of QURA is the utilization of techniques to amplify the amplitude of the desired state without the need for reversed oracles, providing valuable insights into the design of quantum algorithms. 
In particular, our work demonstrates that QURA {achieves a quadratic speedup in computational complexity} over classical methods relying on quantum tomography. As an exact algorithm, QURA eliminates the effect of algorithmic precision on the query complexity of input unitaries, which is unavoidable in classical methods. Hence, our work establishes a promising benchmark for demonstrating meaningful quantum advantage on early fault-tolerant quantum computers.

By offering a deterministic and efficient means of inverting arbitrary unknown unitaries, our work paves the way for the practical implementation of a wide range of quantum algorithms and {provides new research directions in quantum information processing.}
{The quantum circuit architecture in QURA can be constructed via elementary gates, suggesting} its potential to drive the development of meaningful quantum applications, thereby contributing to the ongoing exploration of quantum advantages in early fault-tolerant quantum computing.  
To be applied in practical situations, it is interesting to further consider whether query complexity and computational complexity can be reduced when we know some information of the given unitary, e.g. the multi-qubit evolution has some structure for its Hamiltonian. 
It is also interesting to see whether the complex conjugate of the unitary can be realized more efficiently in this situation, so the dimension of the ancilla system will decrease.

Besides, it is worth noting that our quantum algorithm may offer new perspectives on the physical understanding of time and reversibility at the quantum level. While the arrow of time is a defining characteristic of the macroscopic world, at the quantum scale, reversible transformations are possible if complete knowledge is known. Our quantum method, by circumventing the need for complete information and using finite resources, addresses a fundamental question in quantum physics and illustrates that precise control over unknown quantum evolutions is achievable. This opens new avenues for exploring the frontiers of quantum information processing and developing quantum algorithms that manipulate quantum dynamics. For future directions, it will be interesting to explore how efficiently we can reverse an unknown unitary if we know partial information about the drive Hamiltonian.

\emph{Acknowledgment.}---
Y.-A. C. and Y. M. contributed equally to this work. This work was partially supported by the National Key R\&D Program of China (Grant No.~2024YFE0102500), the National Natural Science Foundation of China (Grant. No.~12447107).
\bibliography{references}

\clearpage
\appendix
\vspace{3cm}
\onecolumngrid

\setcounter{subsection}{0}
\setcounter{table}{0}
\setcounter{figure}{0}

\begin{center}
\Large{\textbf{Supplementary Materials for} \\ \textbf{Quantum Algorithm for Reversing Unknown Unitary Evolutions}}
\end{center}

\numberwithin{equation}{section}
\renewcommand{\theproposition}{S\arabic{proposition}}
\renewcommand{\thedefinition}{S\arabic{definition}}
\renewcommand{\thefigure}{S\arabic{figure}}
\setcounter{equation}{0}
\setcounter{table}{0}
\setcounter{section}{0}
\setcounter{proposition}{0}
\setcounter{definition}{0}
\setcounter{figure}{0}

In this supplemental material, we present detailed proofs of the theorems in the manuscript ``Quantum Algorithm in Reversing Unknown Unitary Evolutions''. 
In Appendix~\ref{appendix:qura}, we present detailed proofs of Theorem 1 in the manuscript that supports our quantum unitary reversal algorithm (QURA). In Appendix~\ref{appendix:qura extension}, we extend our discussions to the cases when the input evolutions are controlled by other systems (\ref{appendix:control gate}), has multiple (qubit) systems (\ref{appendix:multi qubit}), or can be reversed with less slots (\ref{appendix:less slots}).

\begin{table}[ht]
\centering
\setlength{\tabcolsep}{1em}
\caption{A reference of notation conventions in this work.}~\label{tab:notations}
\begin{tabular}{lll}
\toprule
Symbol & Variant & Description \\
\midrule
\addlinespace
$U$ & {} & an unknown unitary transformation  \\
\addlinespace
$d$ & {} & dimension of the quantum system  \\
\addlinespace
$\Delta$ & {$\Delta_d$} & $\arcsin(1/d)$  \\
\addlinespace
$\idop$ & {$\idop_d$} & $d$-dimensional identity operator \\
\addlinespace
$0$ & {} & the number zero or a matrix full of zeros  \\
\addlinespace
$\ket{j}$ & {$\ket{j}_d$} & the $k$-th quantum state from the $d$-dimensional standard basis   \\
\addlinespace
$\omega$ & {} & primitive $d$-th root of unity $e^{2 \pi i / d }$   \\
\addlinespace
$Z$ & {$Z_d$} & clock operator $\sum^{d-1}_{j=0} \omega^{j} \ketbra{j}{j}$ at dimension $d$ \\
\addlinespace
$X$ & {$X_d$} & shift operator $\sum^{d-1}_{j=0} \ketbra{(j + 1) \operatorname{mod} d}{j}$ at dimension $d$ \\
\addlinespace
$\ft$ & {$\ft_d$} & quantum Fourier transformation at dimension $d$ \\
\addlinespace
$\ket{+}$ & {} & a maximal superpositioned state $\sum^{d-1}_{j=0} \ket{j} / \sqrt{d}$ at dimension $d$   \\
\addlinespace
$\ket{00^\perp}$ & {} & normalized state of $\ket{{+}{+}} -\ket{00} / d$\\
\addlinespace
$R_y$ & {} & qubit-Y rotation gate\\
\addlinespace
\midrule
\addlinespace
$\Lambda_0$ & {} & $\ket{00} \ox \idop$, with $\ket{\Lambda_0} = \ket{0}_2 \ox \Lambda_0\ket{\varphi}$\\
\addlinespace
$\Psi_0$ & {} & $\ket{00} \ox U^\dagger$, with $\ket{\Psi_0} = \ket{0}_2 \ox \Psi_0\ket{\varphi}$\\
\addlinespace
$\Lambda_\perp$ & {} & space that is perpendicular to $\Lambda_0$, with  $\ket{\Lambda_\perp} = \ket{0}_2 \ox \Lambda_\perp\ket{\varphi}$\\
\addlinespace
$\Psi_\perp$ & {} & space that is perpendicular to $\Psi_0$, with  $\ket{\Psi_\perp} = \ket{0}_2 \ox \Psi_\perp\ket{\varphi}$\\
\addlinespace
\midrule
\addlinespace
$\cdot$ & {} & matrix multiplication   \\
\addlinespace
$\ket{j k}$ & {$\ket{j, k}$} & $\ket{j} \ox \ket{k}$   \\
\addlinespace
$U^{-1}$ & $U^\dagger$ & unitary inverse  \\
\addlinespace
$U^T$ &  & unitary transpose  \\
\addlinespace
$U^*$ &  & unitary conjugate  \\
\addlinespace
\bottomrule
\end{tabular}
\end{table}
\clearpage

\section{Proof of Main Theorem}~\label{appendix:qura}

In this section, we present the proof of the following theorem stated in the main text:
\renewcommand\theproposition{\ref{thm:unitary inv cir exist}}
\begin{theorem}
For any dimension $d \geq 2$, there exists a quantum algorithm that exactly and deterministically implements $U^\dag$ for arbitrary unitary operator $U \in \su(d)$, by using at most $d \ceil{{\pi}/{2 \Delta}} - 1$ calls of $U$, where $\Delta = \arcsin{{1}/{d}}$.
\end{theorem}
\renewcommand{\theproposition}{S\arabic{proposition}}

\noindent \emph{Sketch of the Proof}.
Here we provide a high-level overview of the proof structure. Detailed constructions of the necessary operators are presented in subsequent subsections.
The central idea is to consider two two-dimensional matrix subspaces $\Span{\Lambda_0, \Lambda_\perp}, \Span{\Psi_0, \Psi_\perp} \subset \CC^{d^3 \times d}$,
with properties
\begin{equation}~\label{eqn:property subspace}
    \Lambda_0 = \ket{00}\ox \idop, \quad 
    \Psi_0 = \ket{00}\ox U^\dagger, \quad 
    \Lambda_\perp^\dagger \Lambda_0 = 0, \quad 
    \Psi_\perp^\dagger \Psi_0 = 0
.\end{equation}
$\Lambda_0$ is the initial state and $\Psi_0$ is the target state.
Explicit choices for $\Lambda_\perp$ and $\Psi_\perp$ are deferred to later subsections.
Next, we construct an \emph{intrinsic encoder} $\encod(U) \in \CC^{d^3 \times d^3}$, which implements $1$ call of the operator $U^*$ and will map the initial state to the target state:
\begin{equation}~\label{eqn:property enc}
    \encod(U)\Lambda_0 = \sin(\Delta)\Psi_0 + \cos(\Delta)\Psi_\perp
.\end{equation}
We then construct an \emph{amplitude amplifier} $\amp{}(U)\in \CC^{2d^3\times 2d^3}$, which uses $1$ call of $U$ and $1$ call of $U^*$. This operator will act as a two-dimensional rotation of a subspace $\mathbb{S} = \Span{\ket{0}_2\ox \Psi_0,\ket{0}_2\ox \Psi_\perp}$,
\begin{equation}~\label{eqn:property amp}
\begin{aligned}
    \amp{}(U) \left(\ket{0}_2\ox \Psi_0 \right) &= \cos(\Delta)\ket{0}_2\ox \Psi_0 - \sin(\Delta) \ket{0}_2\ox \Psi_\perp \textrm{\quad and}\\
    \amp{}(U) \left(\ket{0}_2\ox \Psi_\perp \right) &= \sin(\Delta)\ket{0}_2\ox \Psi_0 + \cos(\Delta) \ket{0}_2\ox \Psi_\perp
.\end{aligned}
\end{equation}
Equivalently, as $\Psi_\perp^\dagger \Psi_0 = 0$, $\amp{}(U)$ can be represented as
\begin{align}
    \amp{}(U) 
    &\equiv \ketbra{0}{0}_2 \ox \left[ \cos(\Delta) \Psi_0 \Psi_0^\dag + \sin(\Delta) \Psi_\perp \Psi_0^\dag - \sin(\Delta) \Psi_0 \Psi_\perp^\dag + \cos(\Delta) \Psi_\perp \Psi_\perp^\dag \right] \\
    &= \begin{bmatrix}
        \cos(\Delta) & \sin(\Delta) \\
        -\sin(\Delta) & \cos(\Delta)
    \end{bmatrix}_{\mathbb{S}} 
    \coloneqq \begin{blockarray}{ccc}
      \ket{0}_2\ox \Psi_0 & \ket{0}_2\ox \Psi_\perp & \\
      \begin{block}{[cc]c}
        \cos(\Delta) & \sin(\Delta) & \bra{0}_2\ox \Psi_0^\dagger \\
        -\sin(\Delta) & \cos(\Delta) & \bra{0}_2\ox \Psi_\perp^\dagger \\
      \end{block}
    \end{blockarray}
.\end{align}
Note that such representation is valid only within the discussion of $\mathbb{S}$. By applying $(m-1)$ repetitions of the amplitude amplifier $\amp{}(U)$ to the encoded initial state, we have
\begin{align}
    \amp{}(U)^{m - 1}\left(\ket{0}_2 \ox \encod(U)\Lambda_0\right)
    &= \amp{}(U)^{m - 1}\left(\sin(\Delta)\ket{0}_2\ox\Psi_0 + \cos(\Delta)\ket{0}_2\ox\Psi_\perp\right)\\
    &= \begin{bmatrix}
        \cos(\Delta) & \sin(\Delta) \\
        -\sin(\Delta) & \cos(\Delta)
    \end{bmatrix}_{\mathbb{S}}^{m - 1} \cdot \begin{bmatrix}
        \sin(\Delta) \\ \cos(\Delta)
    \end{bmatrix}_{\mathbb{S}}  
    = \begin{bmatrix}
        \sin(m\Delta) \\ \cos(m\Delta)
    \end{bmatrix}_{\mathbb{S}} \\
    &= \sin(m\Delta)\ket{0}_2\ox\Psi_0 + \cos(m\Delta)\ket{0}_2\ox\Psi_\perp \label{eqn:approx}
.\end{align}
Consider now an arbitrary $d$-dimensional input state $\ket{\varphi}$. Applying $\idop_{2d^2}\ox \ket{\varphi}$ to both sides of above equation yields:
\begin{equation}
\begin{aligned}
    &\quad\,\, \amp{}(U)^{m - 1}(\idop_2\ox \encod(U))\left(\ket{0}_2\ox \ket{00}\ox \ket{\varphi}\right)\\
    &= \ket{0}_2\ox\left[\sin(m\Delta)\ket{00}\ox U^\dagger\ket{\varphi} + \cos(m\Delta)\Psi_\perp(\ket{00}\ox\ket{\varphi})\right]
.\end{aligned}
\end{equation}
Choosing $m = \pi/(2\Delta)$ exactly gives $\ket{0}_2\ox \ket{00}\ox U^\dagger\ket{\varphi}$. However, since $\pi/(2\Delta)$ generally is not integer-valued for $d>2$, we introduce a variant of the amplitude amplifier, denoted as $\amp{\a}(U)$, which uses the same number of calls to $U$ and $U^*$ as $\amp{}(U)$. This adjusted amplifier will ensure exact and deterministic realization of $U^\dagger$ when we set $m=\ceil{\pi/(2\Delta)} - 1$, i.e.,
\begin{equation}~\label{eqn:property mod amp}
    \amp{\a}(U) \amp{}(U)^{m-1}(\idop_2\ox\encod(U))(\ket{0}_2\ox\ket{00}\ox\ket{\varphi}) 
    = \ket{0}_2\ox\ket{00}\ox U^\dagger\ket{\varphi}
.\end{equation}
Note that Ref.~\cite{Miyazaki2019} has constructed a quantum circuit implementing $U^*$ with $d-1$ calls of $U$ in parallel. 
Counting the total number of calls to $U$, we have $d - 1$ calls from $\encod(U)$ and $(m-1)d$ calls from the repeated amplitude amplification steps, resulting in a total of at most $d\ceil{{\pi}/{2 \Delta}} - 1$ calls. This completes the sketch of the proof.
\hfill$\blacksquare$\vspace{2em}

The remainder of this section verifies the existence and the claimed properties of the sub-circuits $\encod(U)$, $\amp{}(U)$ and $\amp{\a}(U)$, organized as follows.

\begin{enumerate}
    \item [-] Appendix~\ref{appendix:enc} presents the construction of the intrinsic encoder $\encod(U)$ (at Equation~\eqref{eqn:encoder}) that satisfies Equation~\eqref{eqn:property enc}.

    \item [-] Appendix~\ref{appendix:dual} presents the construction of a \emph{shifted decoder} $\decod(U)$ (at Equation~\eqref{eqn:decoder}) as part of substitute for $\encod(U)^\dagger$, and gives the choice of $\Lambda_\perp, \Psi_\perp$ (Equation~\eqref{eqn:subspace}).

    \item [-] Appendix~\ref{appendix:amp} presents the construction of the amplitude amplifier $\amp{}(U)$ (at Equation~\eqref{eqn:amp}) using $\encod(U)$ and $\decod(U)$, that satisfies Equation~\eqref{eqn:property amp}.

    \item [-] Appendix~\ref{appendix:mod amp} presents the construction of the modified amplitude amplifier $\amp{\a}(U)$ (at Equation~\eqref{eqn:mod amp}) that satisfies Equation~\eqref{eqn:property mod amp}.

    \item [-] Appendix~\ref{appendix:qubit} discusses the implementations for the special case $d = 2$.
\end{enumerate}

\subsection{Intrinsic encoder}~\label{appendix:enc}

Recall that the reverse of $U \in \su(d)$ is the transpose conjugate of $U$ as $U^{-1} = U^{\dag}$. 
Here we show $U^\dag$ or $U^T$ can be obtained from $U$ and $U^*$ with following linear equation. 

\begin{lemma}\label{lem:T}
    For any $d$-dimensional unitary $U$, the matrix transpose $U^T$ and conjugate transpose $U^\dag$ can be decomposed into
\begin{align}
     U^T = \dfrac{1}{d} \sum_{j, k = 0}^{d - 1} Z^j X^k U Z^{-j} X^k, \quad 
     U^\dag = \dfrac{1}{d} \sum_{j, k = 0}^{d - 1} Z^j X^k U^* Z^{-j} X^k
,\end{align}
    where $Z = \sum^{d-1}_{j=0} \omega^{j} \ketbra{j}{j}$ with $\omega = e^{2 \pi i / d }$, and $X = \sum^{d-1}_{j=0} \ketbra{(j + 1) \operatorname{mod} d}{j}$ are the clock and shift operators of dimension $d$, respectively.
\end{lemma}
\begin{proof}
Since such two decompositions are equivalent, it is enough to prove only the first one.
For fixed computational basis $\ket{m}$ and $\ket{n}$ in the $d$-dimensional Hilbert space, it is checked that
\begin{align}
    &\braandket{m}{\sum_{j, k = 0}^{d - 1} Z^j X^k U Z^{-j} X^k}{n} 
    = \sum_{j, k = 0}^{d - 1} \braandket{m}{Z^j X^k U Z^{-j} X^k}{n}\\
    =& \sum_{j, k} \omega^{-j (-m + n + k)} \braandket{m - k}{U}{n + k}
    =\sum_{k} \braandket{m - k}{U}{n + k}\cdot\sum_j\omega^{-j (-m + n + k)}\\
    =& \braandket{m - (m-n)}{U}{n + (m-n)}\cdot d
    = d \braandket{n}{U}{m} = d \braandket{m}{U^T}{n}.
\end{align}
\end{proof}

\begin{remark}
 Lemma~\ref{lem:T} reveals a linear decomposition of matrix transposition. One may wonder whether such a decomposition exists for the conjugate transpose. We remark that conjugate transpose is not a $\CC$-linear operator, which implies such linear decomposition does not exist.
\end{remark}

Inspired by Lemma~\ref{lem:T} and the LCU algorithm~\cite{childs2012hamiltonian}, one could construct a select gate
\begin{equation}~\label{eqn:app-S2-EFT}
    \encod_{\ft}(U) \coloneqq \sum_{j,k=0}^{d-1}\ketbra{j,k}{j,k}\ox Z^j X^k U Z^{-j} X^k
\end{equation}
and the intrinsic encoder
\begin{equation}~\label{eqn:encoder}
    \encod(U) \coloneqq \left(\ft^{\dag} \ox \ft^\dag \ox  \idop\right)\cdot \encod_{\ft}(U^*)\cdot \left(\ft^{\dag} \ox \ft^\dag \ox  \idop\right)
,\end{equation}
where $\ft = \sum_{j,k=0}^{d-1}\omega^{jk}\ketbra{j}{k} / \sqrt{d}$ is the $d$-dimensional Fourier transform. The select gate $\encod_{\ft}(U^*)$ could be implemented by $2d$ control-$Z^j$ gates, $2d$ control-$X^k$ gates and $1$ call of $U^*$.

\begin{lemma}~\label{lem:preparation}
    Let $\Lambda_0, \Psi_0$ be as defined in Equation~\eqref{eqn:property subspace}. Then
\begin{equation}
    \encod(U)\Lambda_0 = \sin(\Delta)\Psi_0 + \cos(\Delta)\Psi_\perp
,\end{equation}
    where $\Psi_\perp \in \CC^{d^3\times d}$ satisfies $\Psi_\perp^\dagger \Psi_0 = 0$.
\end{lemma}
\begin{proof}
    This is done by a direct computation.
\begin{align}
    \encod(U)\Lambda_0 &= \left(\ft^{\dag} \ox \ft^\dag \ox  \idop\right) \encod_{\ft}(U^*) \left(\ft^{\dag} \ox \ft^\dag \ox  \idop\right) \cdot \left( \ket{00} \ox \idop \right) \\
    &= \frac{1}{d} \left(\ft^{\dag} \ox \ft^\dag \ox  \idop\right)
    \left(\sum_{j,k=0}^{d-1}\ketbra{j,k}{j,k}\ox Z^j X^k U^* Z^{-j} X^k\right) \cdot
    \sum_{j,k=0}^{d-1} \ket{j, k} \ox \idop \\
    &= \frac{1}{d} \left(\ft^{\dag} \ox \ft^\dag \ox  \idop\right)
    \cdot \sum_{j,k=0}^{d-1}\ket{j,k}\ox Z^j X^k U^* Z^{-j} X^k \\
    &= \frac{1}{d^2} \left(\sum_{r, t, j, k = 0}^{d - 1} \omega^{-rj-tk} \ketbra{r, t}{j, k} \ox \idop\right)
    \cdot \sum_{j,k=0}^{d-1}\ket{j,k}\ox Z^j X^k U^* Z^{-j} X^k \\
    &= \frac{1}{d^2} \sum_{r, t, j, k = 0}^{d - 1} \omega^{-rj-tk} \ket{r, t} \ox Z^j X^k U^* Z^{-j} X^k
.\end{align}
Consider
\begin{equation}
    \left(\ketbra{00}{00}\ox \idop\right)\cdot\encod(U)\Lambda_0= \frac{1}{d^2}\ket{00}\ox \sum_{j, k = 0}^{d - 1} Z^j X^k U^* Z^{-j} X^k=\frac{1}{d}\ket{00}\ox U^\dag=\sin(\Delta)\Psi_0,
\end{equation}
we obtain
\begin{equation}
    \encod(U)\Lambda_0 = \sin(\Delta)\Psi_0 + \cos(\Delta)\Psi_\perp
\text{, where }
\Psi_\perp= \frac{1}{\cos(\Delta)}\left(\left(\idop_{d^2}-\ketbra{00}{00}\right)\ox \idop\right)\encod(U)\Lambda_0
.\end{equation}
This completes the proof as $\Psi_\perp^\dag\Psi_0 = 0$ by construction.
As a side note, Equation~\eqref{eqn:init} will give a more explicit form of $\Psi_\perp$.
\end{proof}

\subsection{Duality relations}~\label{appendix:dual}

The shifted decoder is constructed as
\begin{equation}~\label{eqn:decoder}
    \decod(U) \coloneqq \sum_{j,k=0}^{d-1}\ketbra{j,k}{j,k}\ox X^{-j}Z^{-k}UX^{-j}Z^k
.\end{equation}
Then one can derive a duality relation between $\decod(U)$ and 
$\encod_{\ft}$.

\begin{lemma}[Dual relations]\label{lem:Q1Q2}  
For $\ket{+} = \sum_{j=0}^{d-1}\ket{j} / \sqrt{d}$,
\begin{align}
    &(\ft^{\dag} \ox \ft^\dag \ox \idop)\cdot \encod_{\ft}(U^*)\cdot(\ket{{+}{+}} \ox\idop) = \decod(U)^\dag\cdot(\ket{{+}{+}} \ox\idop), \label{eqn:duality_1}\\
    &(\ft^{\dag} \ox \ft^\dag \ox  \idop)\cdot \decod(U)\cdot(\ket{{+}{+}} \ox\idop)=\encod_{\ft}(U^*)^\dag\cdot(\ket{{+}{+}} \ox\idop)\label{eqn:duality_2}
.\end{align}
\end{lemma}
\begin{proof}
By introducing the clock and shift operators, we denote the decomposition of $U^*$ on Weyl operator basis as
\begin{equation}\label{eqn:U_decom}
    U^* = \sum_{\a,\beta=0}^{d-1}p_{\a,\beta}Z^\a X^\beta,
\end{equation}
and then we have
\begin{equation}
    \begin{aligned}
        U=\sum_{\a,\beta=0}^{d-1}p_{\a,\beta}^*Z^{-\a} X^\beta,\ 
        U^T=\sum_{\a,\beta=0}^{d-1}p_{\a,\beta}^* X^{-\beta}Z^{-\a}.\ 
        U^\dag=\sum_{\a,\beta=0}^{d-1}p_{\a,\beta} X^{-\beta}Z^\a.
    \end{aligned}
\end{equation}
For any $0 \leq r, t < d$, we find 
\begin{align}
    &(\bra{r, t}\ox\idop)\cdot(\ft^{\dag} \ox \ft^\dag \ox  \idop)\cdot \encod_{\ft}(U^*)\cdot(\ket{{+}{+}} \ox\idop)\\
    =&(\bra{r, t}\ox\idop)\cdot(\ft^{\dag} \ox \ft^\dag \ox  \idop)\cdot\frac{1}{d}\sum_{j,k=0}^{d-1}\ket{j, k}\ox Z^jX^kU^{*}Z^{-j}X^k\\
    =&\frac{1}{d^2}\sum_{j,k=0}^{d-1}\omega^{-rj-tk} Z^jX^k\cdot\sum_{\a,\beta=0}^{d-1}p_{\a,\beta}Z^\a X^\beta\cdot Z^{-j}X^k\\
    =&\frac{1}{d^2}\sum_{\a,\beta=0}^{d-1}p_{\a,\beta}\sum_{j,k=0}^{d-1}\omega^{j(-r+k+\beta)-(t+\a) k} Z^\a X^{\beta+2k}\\
    =&\frac{1}{d^2}\sum_{\a,\beta=0}^{d-1}p_{\a,\beta}\sum_{k=0}^{d-1}\omega^{-(t+\a)k} Z^\a X^{\beta+2k}\cdot\sum_{j=0}^{d-1}\omega^{j(-r+k+\beta)}\\
    =&\frac{1}{d^2}\sum_{\a,\beta=0}^{d-1}p_{\a,\beta}\cdot\omega^{-(t+\a)(r-\beta)} Z^\a X^{\beta+2(r-\beta)}\cdot d\\
    =&\frac{1}{d}\sum_{\a,\beta=0}^{d-1}p_{\a,\beta}Z^{-t}X^{r} X^{-\beta}Z^\a Z^{t}X^r\\
    =&\frac{1}{d} \ Z^{-t}X^{r}U^\dag Z^{t}X^r\\
    =&\frac{1}{d} \ (X^{-r}Z^{-t}UX^{-r}Z^t)^\dag\\
    =&(\bra{r, t}\ox\idop)\cdot \decod(U)^\dag\cdot(\ket{{+}{+}} \ox\idop).
\end{align}
Similarly, another relation takes
\begin{align}
    &(\bra{r, t}\ox\idop)\cdot(\ft^{\dag} \ox \ft^\dag \ox  \idop)\cdot \decod(U)(\ket{{+}{+}} \ox\idop)\\
    =&(\bra{r, t}\ox\idop)\cdot(\ft^{\dag} \ox \ft^\dag \ox  \idop)\cdot\frac1d\sum_{j,k=0}^{d-1}\ket{j, k}\ox X^{-j}Z^{-k}UX^{-j}Z^k\\ 
    =&\frac{1}{d^2}\sum_{j,k=0}^{d-1}\omega^{-rj-tk} X^{-j}Z^{-k}\cdot\sum_{\a,\beta=0}^{d-1}p_{\a,\beta}^*Z^{-\a} X^\beta \cdot X^{-j}Z^k\\
    =&\frac{1}{d^2}\sum_{\a,\beta=0}^{d-1}p_{\a,\beta}^*\sum_{j,k=0}^{d-1}\omega^{-rj-tk-(k+\a)(\beta-j)} X^{\beta-2j}Z^{-\a}\\
    =&\frac{1}{d^2}\sum_{\a,\beta=0}^{d-1}p_{\a,\beta}^*\sum_{j}^{d-1}\omega^{-rj-\a(\beta-j)} X^{\beta-2j}Z^{-\a}\cdot\sum_{k=0}^{d-1}\omega^{-k(t+\beta-j)}\\
    =&\frac{1}{d^2}\sum_{\a,\beta=0}^{d-1}p_{\a,\beta}^*\cdot\omega^{-r(t+\beta)-\a(\beta-(t+\beta))} X^{\beta-2(t+\beta)}Z^{-\a}\cdot d\\    
    =&\frac1d \sum_{\a,\beta=0}^{d-1}p_{\a,\beta}^* X^{-t}Z^{r}X^{-\beta}Z^{-\a} X^{-t}Z^{-r}\\
    =&\frac1d X^{-t}Z^{r}U^T X^{-t}Z^{-r}\\
    =&\frac1d (Z^rX^tU^*Z^{-r}X^t)^\dag\\
    =&(\bra{r, t}\ox\idop)\cdot \encod_{\ft}(U^*)^\dag\cdot(\ket{{+}{+}} \ox\idop),
\end{align}
which suggests the dual relations hold. 
\end{proof}

\begin{remark}
    Either Equation~\eqref{eqn:duality_1} or Equation~\eqref{eqn:duality_2} could be considered as a definition equation of $\decod(U)$ with assumption that $\decod(U)$ is a block-diagonal matrix, i.e. of form         $\sum_{j,k=0}^{d-1}\ketbra{j,k}{j,k}\ox V_{jk}$ with each $V_{jk}\in\mathbb C^{d\times d}$. That is, if $\decod(U)$ is block-diagonal and defined by Equation~\eqref{eqn:duality_2}, then $\decod(U)$ will satisfy Equation~\eqref{eqn:decoder} and Equation~\eqref{eqn:duality_1}.
\end{remark}

\begin{corollary}[Dual relations, simplified]\label{cor:O1}
\begin{align}
    &\encod(U)\Lambda_0 = \decod(U)^\dag (\ket{{+}{+}} \ox\idop), \label{eqn:s_dual_1}\\
    &\decod(U) (\ket{{+}{+}} \ox \idop) = \encod(U)^\dag\Lambda_0 \label{eqn:s_dual_2}.
\end{align}
\end{corollary}
\begin{proof}
By Lemma~\ref{lem:Q1Q2}, we rewrite Equation~\eqref{eqn:duality_1} as
\begin{align}
    &(\ft^{\dag} \ox \ft^\dag \ox  \idop)\cdot \encod_{\ft}(U^*)\cdot(\ft^{\dag} \ox \ft^\dag \ox  \idop)\cdot(\ket{00} \ox\idop)\\
    =&(\ft^{\dag} \ox \ft^\dag \ox  \idop)\cdot \encod_{\ft}(U^*)\cdot(\ket{{+}{+}} \ox\idop)\\
    =&\decod(U)^\dag\cdot(\ket{{+}{+}} \ox\idop),
\end{align}
and apply $(\ft \ox \ft \ox  \idop)$ on Equation~\eqref{eqn:duality_2} 
\begin{align}
     \decod(U)\cdot(\ket{{+}{+}} \ox\idop)
    =&(\ft  \ox \ft  \ox  \idop)\cdot \encod_{\ft}(U^*)^\dag \cdot(\ket{{+}{+}} \ox\idop)\\
    =& \encod(U)^\dag \cdot(\ket{00} \ox\idop).
\end{align}
\end{proof}
\begin{remark}\label{remark 4}
By marking two orthogonal components $\ket{00} $ and $\ket{00^\perp}$ in $d^2$-dimensional qudit space, $\ket{{+}{+}} $ can be decomposed into
\begin{equation}\label{eqn:decom}
    \ket{{+}{+}} = \sin(\Delta) \ket{00}  + \cos(\Delta) \ket{00^\perp},
\end{equation}
wher $\ket{00^\perp}\coloneqq \left(\ket{{+}{+}} -\sin(\Delta)\ket{00} \right)/\cos(\Delta)$. Therefore, Equation~\eqref{eqn:s_dual_1} can be seen as an initializing equation
\begin{equation}~\label{eqn:init}
    \encod(U) \Lambda_0
    = \sin(\Delta) \decod(U)^\dag \Lambda_0 
    + \cos(\Delta) \decod(U)^\dag (\ket{00^{\perp}} \ox \idop)
.\end{equation}
Observe that
\begin{equation}~\label{eqn:D_dag_0}
    \decod(U)^\dag \Lambda_0 = \decod(U)^\dag (\ket{00} \ox \idop)
    = \ket{00} \ox U^\dag = \Psi_0
,\end{equation}
which gives a simplified form of $\Psi_\perp$ in Lemma~\ref{lem:preparation}.
\end{remark}

Corollary~\ref{cor:O1} not only reveals an initialization for obtaining unitary inverse, but also implicitly suggests a map between two linear spaces, which is critical to amplify the amplitude of the desired component.
By far, we can formally define two subspaces introduced in Equation~\eqref{eqn:property subspace}:
\begin{equation}~\label{eqn:subspace}
\begin{aligned}
     \Lambda_0 &= \ket{00}\ox \idop, &&\Lambda_\perp = \decod(U) \left(\ket{00^\perp} \ox U^\dag\right); \\
     \Psi_0 &= \ket{00}\ox U^\dag, &&\Psi_\perp = \decod(U)^\dagger \left(\ket{00^\perp} \ox \idop \right)
.\end{aligned}
\end{equation}
Then one could construct the operations between two subspaces $\Span{\Lambda_0, \Lambda_\perp}, \Span{\Psi_0, \Psi_\perp}$ with the help of an additional qubit zero state, as shown in the next subsection.

\subsection{Duality-based amplitude amplifier}~\label{appendix:amp}

$\encod(U)$ can be understood as a rotation between $\Span{\Lambda_0, \Lambda_\perp}, \Span{\Psi_0, \Psi_\perp}$, i.e.,
\begin{equation}
    \encod(U)
    = \begin{blockarray}{ccc}
      \Psi_0 & \Psi_\perp & \\
      \begin{block}{[cc]c}
        \sin(\Delta) & \cos(\Delta) & \Lambda_0^\dagger \\
        \cos(\Delta) & -\sin(\Delta) & \Lambda_\perp^\dagger \\
      \end{block}
    \end{blockarray}
\end{equation}
as proved by the following lemma.

\begin{lemma}\label{lem:Lambda_perp}
For $\Lambda_0, \Lambda_\perp, \Psi_0, \Psi_\perp$ defined in Equation~\eqref{eqn:subspace}, $\encod(U)$ satisfies
\begin{equation}~\label{eqn:E_Lambda}
\begin{aligned}
    \encod(U) \Lambda_{0} &= \sin(\Delta) \Psi_0 + \cos(\Delta) \Psi_\perp, \\
    \encod(U) \Lambda_{\perp} &= \cos(\Delta) \Psi_0 - \sin(\Delta) \Psi_\perp.
\end{aligned}
\end{equation}
\end{lemma}
\begin{proof}
The first statement directly follows by Equation~\eqref{eqn:init} and Equation~\eqref{eqn:D_dag_0}. 
For the second one,
following the definition of $\Lambda_\perp$, we find 
\begin{align}
    \Lambda_\perp
    =&\,\, \decod(U) (\ket{00^\perp} \ox U^\dag)
    = \decod(U) (\ket{00^\perp} \ox \idop) \cdot (\idop_{d^2} \ox U^\dag) \\
    =&\,\, \frac{1}{\sqrt{d^2-1}} \left[ d \cdot \decod(U) (\ket{{+}{+}} \ox \idop)- \decod(U) (\ket{00} \ox \idop) \right] \cdot (\idop_{d^2} \ox U^\dag)\label{eqn:s49}\\
    \xlongequal{\eqref{eqn:s_dual_2},\eqref{eqn:decoder}}&
    \frac{1}{\sqrt{d^2-1}} \left[ d \cdot \encod(U)^\dag (\ket{00} \ox \idop) - \ket{00} \ox U \right] \cdot (\idop_{d^2} \ox U^\dag)\\
    =&\,\, \frac{1}{\sqrt{d^2-1}} \encod(U)^\dag \cdot \left[ {d\ket{00} \ox U^\dag-\encod(U) (\ket{00} \ox\idop)} \right]\\
    \xlongequal{\eqref{eqn:D_dag_0},\eqref{eqn:s_dual_1}}& \frac{1}{\sqrt{d^2-1}}  \encod(U)^\dag \cdot \left[ {d\decod(U)^\dag (\ket{00} \ox \idop)-\decod(U)^\dag  (\ket{{+}{+}} \ox \idop)} \right]\\   
    =&\,\, \encod(U)^\dag \cdot \frac{1}{\sqrt{d^2-1}} \decod(U)^\dag \left[ (d\ket{00} -\ket{{+}{+}}) \ox \idop \right]\\
    =&\,\, \encod(U)^\dag \cdot \frac{1}{d} \decod(U)^\dag \left((\sqrt{d^2-1}\ket{00} -\ket{00^\perp}) \ox\idop\right)\label{eqn:s54}\\    
    =&\,\, \encod(U)^\dag \cdot \left(\cos(\Delta) \Psi_0 - \sin(\Delta) \Psi_\perp\right)
\end{align}
The result follows by applying $\encod(U)$ on both sides.
\end{proof}
\vspace{2em}

Apart from the trivial inverse map $\encod^\dag(U)$, we may consider another unitary $\widetilde{\decod}(U)$ mapping $\Span{\Psi_0,\Psi_\perp}$ back into $\Span{\Lambda_0,\Lambda_\perp}$.
Such an inverse map can be constructed with the help of an ancillary qubit and two fixed quantum gate $G$. Denote $X_2$ as the Pauli-X gate on the ancillary qubit system, and let $F$ be any $d^2$-dimensional unitary satisfying $F \ket{00} = \ket{00^\perp}$. Then quantum gate $G$ and unitary map $\widetilde{\decod}(U)$ on the qubit-qudit system are defined as
\begin{align}\label{eqn:G and D}
    G  &\coloneqq (\ketbra{1}{1} \ox F + \ketbra{0}{0} \ox \idop)\cdot((X_2 - \idop_2) \ox \ketbra{00}{00} + \idop_2 \ox \idop ), \\
     \widetilde{\decod}(U) &\coloneqq (G^\dag\ox\idop)\cdot(X_2\ox \decod(U))\cdot(G\ox\idop).
\end{align}
In the following statement, $\widetilde{\decod}(U)$ will be proved to map $\Psi_\perp$ into $\Lambda_0$, and $\Psi_0$ into $\Lambda_\perp$, i.e.,
\begin{equation}
    \widetilde{\decod}(U) = 
    \begin{blockarray}{ccc}
      \ket{0}_2 \ox \Lambda_0 & \ket{0}_2 \ox \Lambda_\perp & \\
      \begin{block}{[cc]c}
        0 & 1 & \bra{0}_2 \ox \Psi_0^\dagger \\
        1 & 0 & \bra{0}_2 \ox \Psi_\perp^\dagger \\
      \end{block}
    \end{blockarray}
.\end{equation}
Notice that the construction of $\widetilde{\decod}(U)$ with an additional qubit system is nontrivial since $\decod(U)^\dag$ is not available in the absence of $U^\dag$.

\begin{figure}[t]
\captionsetup{justification=raggedright, singlelinecheck=false}
\centering
\includegraphics[width=0.8\linewidth]{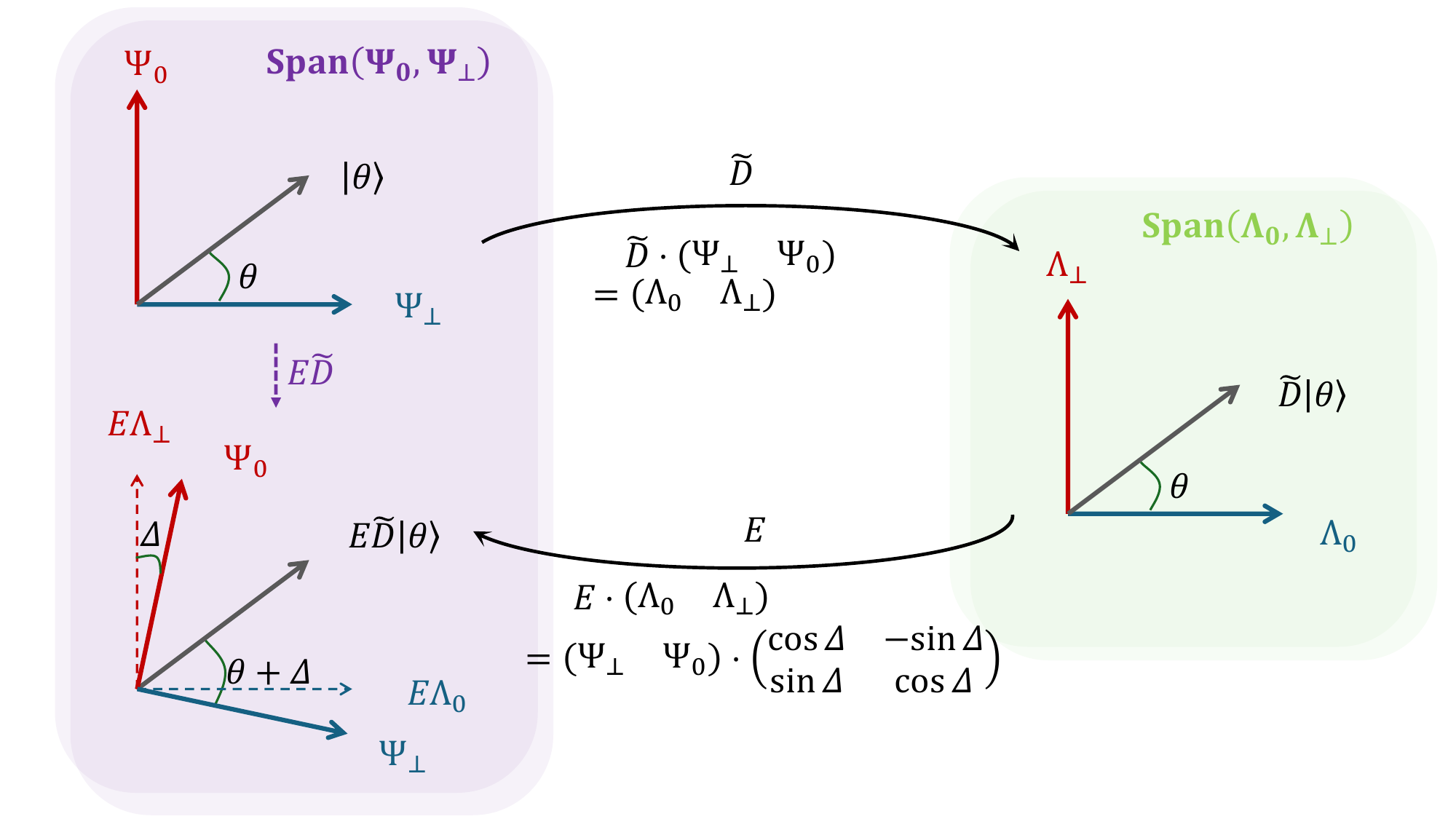}
\caption{Depiction of the rotations between the two dual subspaces, where the integrated rotation $E\Tilde{D}$ achieves a $\Delta$ angle amplifying to the target $\Psi_0$.}
\end{figure}

\begin{lemma}\label{lem:tilde_D}
    Following the definitions above, we have
\begin{equation}~\label{eqn:tilde_D}
\begin{aligned}
    \widetilde{\decod}(U) ( \ket{0}_2 \ox\Psi_0) &= \ket{0}_2 \ox \Lambda_\perp, \\
    \widetilde{\decod}(U) ( \ket{0}_2 \ox\Psi_\perp) &= \ket{0}_2 \ox \Lambda_0
.\end{aligned}
\end{equation}
\end{lemma}
\begin{proof}
The intermediate steps of applying $\widetilde{\decod}(U)$ on $\ket{0} \ox \Psi_0$ and $\ket{0} \ox \Psi_\perp$ are 
\begin{enumerate}
    \item [-] $\ket{0}_2 \ox \Psi_0 
    \xrightarrow{G\ox\idop} \ket{1}_2 \ox \ket{00^\perp} \ox U^\dag 
    \xrightarrow{X\ox \decod(U)} \ket{0}_2 \ox \Lambda_\perp \xrightarrow{G^\dag\ox\idop} \ket{0}_2 \ox \Lambda_\perp$, and
    \item [-] $\ket{0}_2 \ox \Psi_\perp
    \xrightarrow{G\ox\idop }  
    \ket{0}_2 \ox \decod(U)^\dag (\ket{00^\perp} \ox\idop)  
    \xrightarrow{X\ox \decod(U)}\ket{1}_2 \ox \ket{00^\perp} \ox\idop
    \xrightarrow{G^\dag\ox\idop }
    \ket{0}_2 \ox \Lambda_0$.
\end{enumerate}
\end{proof}

By combining Lemma~\ref{lem:Lambda_perp} and Lemma~\ref{lem:tilde_D}, we are ready to construct the amplitude amplifier $\amp{}(U)$ as
\begin{equation}~\label{eqn:amp}
    \amp{}(U) \coloneqq (\idop_2 \ox \encod(U)) \cdot \widetilde{\decod}(U)
.\end{equation}
Since such construction is based on the duality relation, we call $\amp{}(U)$ the \emph{duality-based amplitude amplifier}.
As described in Figure~\ref{fig:duality}, the amplifier is essentially a rotation on the subspace $\Span{\ket{0}_2 \otimes\Psi_0,\ket{0}_2 \otimes\Psi_\perp}$, given by the following lemma.

\begin{figure}[t]
\captionsetup{justification=raggedright, singlelinecheck=false}
\centering
\includegraphics[width=0.8\linewidth]{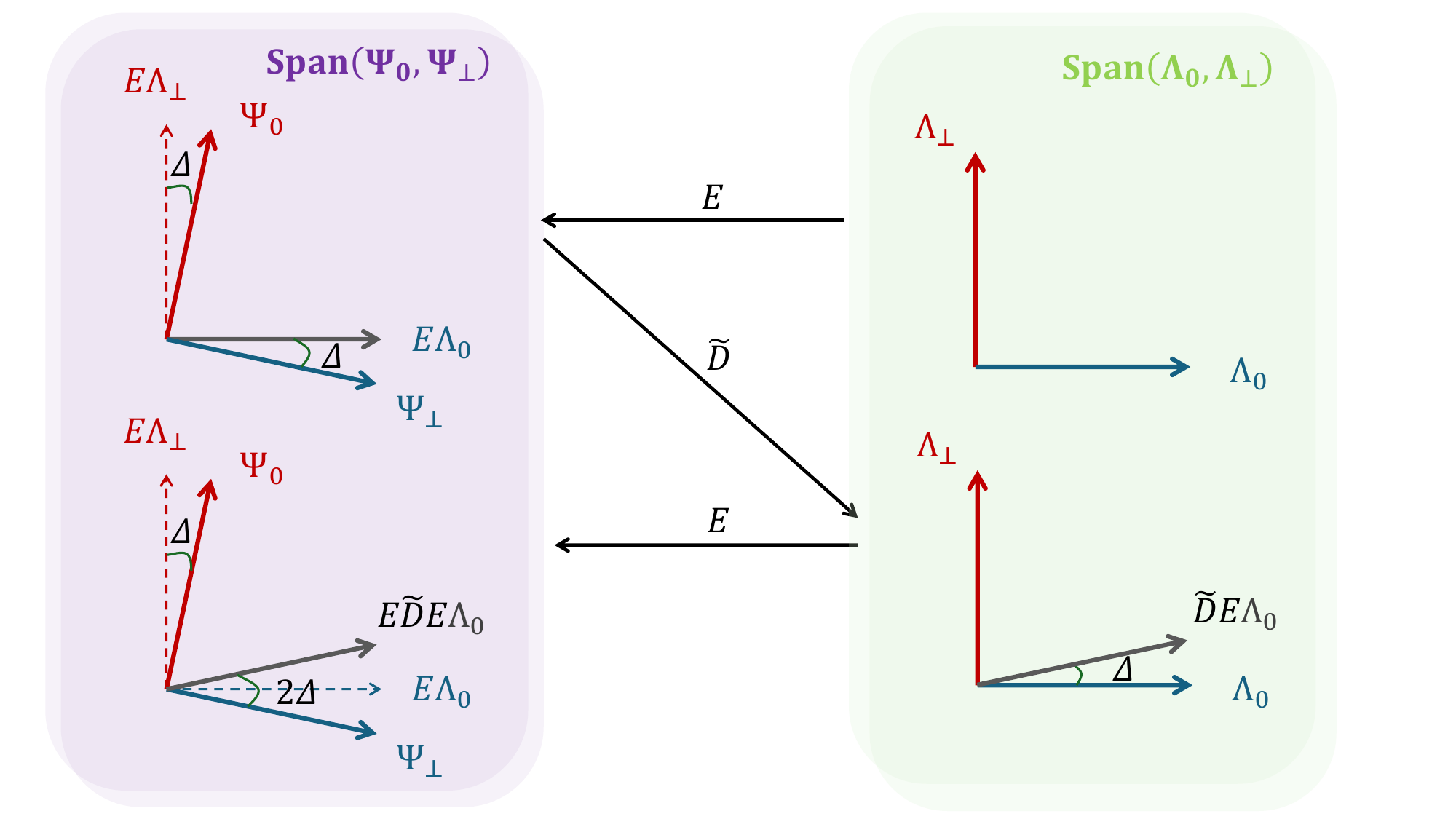}
\caption{Depiction of Initialization and first round of amplitude amplify. The state starts from $ \Lambda_0 $ and after $\encod \cdot \widetilde{\decod} \cdot \encod$, we reach the state $\encod \widetilde{\decod} \encod  \Lambda_0 $ with $\sin(2\Delta)$ amplitude component of the target state $\Psi_0$.}
\label{fig:duality 2}
\end{figure}

\begin{lemma}\label{lem:oracle_d}
Following the definitions above, we have
\begin{equation}
\begin{aligned}
    \amp{}(U) \left(\ket{0}_2\ox \Psi_0 \right) &= \cos(\Delta)\ket{0}_2\ox \Psi_0 - \sin(\Delta) \ket{0}_2\ox \Psi_\perp \textrm{\quad and}\\
    \amp{}(U) \left(\ket{0}_2\ox \Psi_\perp \right) &= \sin(\Delta)\ket{0}_2\ox \Psi_0 + \cos(\Delta) \ket{0}_2\ox \Psi_\perp
,\end{aligned}
\end{equation}
\end{lemma}
\begin{proof}
This statement is a direct implication of Lemma~\ref{lem:Lambda_perp}, Lemma~\ref{lem:tilde_D} and the construction of $\amp{}(U)$:
\begin{align}
    \amp{}(U) \left(\ket{0}_2\ox \Psi_0 \right) 
    &= (\idop_2 \ox \encod(U)) \widetilde{\decod}(U) \cdot \left(\ket{0}_2\ox \Psi_0 \right) \\
    &= (\idop_2 \ox \encod(U)) \cdot \left(\ket{0}_2\ox \Lambda_\perp \right) \\
    &= \ket{0}_2 \ox \left( \cos(\Delta) \Psi_0 - \sin(\Delta) \Psi_\perp \right) \\
    &= \cos(\Delta)\ket{0}_2\ox \Psi_0 - \sin(\Delta) \ket{0}_2\ox \Psi_\perp
.\end{align}
Similar reasoning hold for the second statement:
\begin{align}
    \amp{}(U) \left(\ket{0}_2\ox \Psi_\perp \right) 
    &= (\idop_2 \ox \encod(U)) \cdot \left(\ket{0}_2\ox \Lambda_0 \right) \\
    &= \sin(\Delta)\ket{0}_2\ox \Psi_0 + \cos(\Delta) \ket{0}_2\ox \Psi_\perp
.\end{align}
\end{proof}

{Figure~\ref{fig:duality 2} shows the complete process to construct $\sin(2\Delta)$ component of the target $\Psi_0$, which is equivalent to implement the Initialization and first round of $\amp{}(U)$.}

\subsection{Modified amplitude amplifier}~\label{appendix:mod amp}

\begin{figure}[H]
\captionsetup{justification=raggedright, singlelinecheck=false}
\centering
\includegraphics[width=0.8\linewidth]{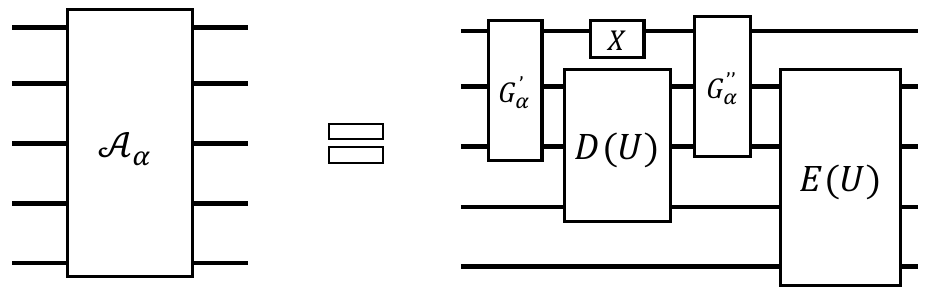}
\caption{Quantum circuit representation of the amplifier $\cA_{\a}$ with respect to the definition in Eq~\eqref{eqn:mod amp}. }
\label{fig: QURA Amplifier1}
\end{figure}

To get an exact and deterministic implementation for $U^\dag$, we need to tune $\amp{}(U)$ to realize amplitude amplification of consecutive angles. Amplifying a given angle requires knowledge of $\theta$, which is met when the dimension $d$ is specified in our task. Let $F_\a \in \CC^{d^2 \times d^2}$ be any unitary satisfying
\begin{equation}\label{eqn:Fa}
    F_\a \ket{00} =\sqrt{1-\a^2} \ket{00} +\a \ket{00^\perp},
\end{equation}
and let
\begin{align}
    G_\a^{'} &\coloneqq (\ketbra{1}{1}_2 \ox F_\a + \ketbra{0}{0}_2 \ox \idop) \cdot ( (X-\idop_2)\ox \ketbra{00}{00}_2 +\idop_2 \ox \idop),\\
    G_{\a}^{''} &\coloneqq \left((R_y(t)-\idop_2)\ox\ketbra{00}{00} +\idop_2 \ox \idop\right) \cdot (\ketbra{1}{1}_2 \ox F^\dag + \ketbra{0}{0}_2 \ox\idop)
,\end{align}
where $R_y(t)$ denotes a single-qubit {$Y$-}rotation of angle $-2\arctan\frac{1}{\tan{\theta}\sqrt{1-\a^2}}$.
Then the modified amplitude amplifier is constructed as
\begin{equation}~\label{eqn:mod amp}
    \amp{\a,\theta}(U) \coloneqq (\idop_2 \ox \encod(U))\cdot(G_\a^{''}\ox\idop)\cdot(X\ox \decod(U))\cdot(G_\a^{'}\ox\idop)
.\end{equation}
We depict the construction of the amplifier $\amp{\a,\theta}(U)$ in Figure~\ref{fig: QURA Amplifier1}, where $\amp{\a,\theta}(U)$ is abbreviated as $\amp{\a}(U)$.

\begin{lemma}\label{lem:mod}
Let $\theta\in[0,2\pi), \a\in[0,1]$. Then
\begin{equation}
    \amp{\a,\theta}(U) \left(\sin(\theta)  \ket{0}_2 \ox \Psi_0 + \cos(\theta)  \ket{0}_2 \ox\Psi_\perp\right)
    = \sin(\theta') \ket{0}_2 \ox\Psi_0 + \cos(\theta') \ket{0}_2 \ox\Psi_\perp
,\end{equation}
where $\theta'=\arcsin(\a \sin(\theta))+\Delta$.
\end{lemma}
\begin{proof}
An initial state in the space of $\{\Psi_0, \Psi_\perp\}$ is transformed by the amplifier as
\begin{align}
    &\,\, \sin(\theta) \ket{0} \ox  \Psi_0 + \cos(\theta)\ket{0} \ox  \Psi_\perp\\
    =&\,\, \sin(\theta)  \ket{0}_2 \ox \ket{00} \ox U^\dag+ \cos(\theta)\ket{0} \ox \Psi_\perp\\
    \xrightarrow{(X-\idop_2)\ox\ketbra{00}{00} \ox\idop+\idop}
    &\,\,  \sin(\theta)  \ket{1}_2 \ox \ket{00} \ox U^\dag + \cos(\theta)\ket{0} \ox \Psi_\perp\\
    \xrightarrow{\ketbra{1}{1}\ox F_\a\ox\idop+\ketbra{0}{0}\ox\idop}
    &\,\,  \sin(\theta) \ket{1}_2 \ox\left(\sqrt{1-\a^2} \ket{00} +\a \ket{00^\perp}\right)\ox U^\dag + \cos(\theta)\ket{0} \ox \Psi_\perp\\
    \xrightarrow{X\otimes \decod(U)}
    &\,\,  \sin(\theta) \ket{0}_2 \ox\left(\sqrt{1-\a^2} \Lambda_0 +\a \Lambda_\perp \right)
    + \cos(\theta) \ket{1}_2 \ox \ket{00^\perp}\ox\idop\\
    \xrightarrow{\ketbra{1}{1}\ox F^\dag\ox\idop+\ketbra{0}{0}\ox\idop}
    &\,\,  \sin(\theta) \ket{0}_2 \ox\left(\sqrt{1-\a^2} \Lambda_0 +\a \Lambda_\perp \right) + \cos(\theta) \ket{1}_2 \ox \Lambda_0 \\
    =&\,\,  \left(\sqrt{1-\a^2}\sin(\theta) \ket{0}_2 + \cos(\theta)\ket{1}\right) \Lambda_0 +\a\sin(\theta) \ket{0}_2 \ox \Lambda_\perp 
    \\
    \xrightarrow{(Ry(t)-\idop_2)\ox\ketbra{00}{00} \ox\idop+\idop_2 \ox \idop}
    &\,\,  \ket{0}_2 \ox \left(\sqrt{1-\a^2\sin^2\theta} \Lambda_0 + \a\sin(\theta) \Lambda_\perp\right)
.\end{align}
By Lemma~\ref{lem:Lambda_perp}, observe that
\begin{align}
   &\,\, \encod(U) \Lambda_0 \left(\sqrt{1-\a^2\sin^2\theta} \Lambda_0 + \a\sin(\theta) \Lambda_\perp\right) \\
   =&\,\, \left( \sqrt{1-\a^2\sin^2\theta} \sin(\Delta)  + \a \sin(\theta) \cos(\Delta) \right) \Psi_0 + \\
  &\,\,\left( \sqrt{1-\a^2\sin^2\theta} \cos(\Delta)  - \a \sin(\theta) \sin(\Delta) \right)\Psi_\perp \\
  =&\,\, \sin\left(\arcsin(\a\sin\theta)+\Delta\right) \Psi_0 + \cos \left(\arcsin(\a\sin\theta)+\Delta\right) \Psi_\perp
\end{align}
Therefore, applying $\idop_2\ox\encod(U)$ in the end gives the desired result.
\end{proof}
\vspace{1em}

\begin{remark}
    In Lemma~\ref{lem:mod}:
\begin{enumerate}
    \item [-] since it is easy to transform $\sin(\theta) \Psi_0 + e^{i\gamma}\cos(\theta)\Psi_\perp$ into $\sin(\theta) \Psi_0 + \cos(\theta)\Psi_\perp$ when $\gamma$ known, we could change the input state to $\sin(\theta) \Psi_0 + e^{i\gamma}\cos(\theta)\Psi_\perp$;
    \item [-] when we set $\a=1$, gates $F_\a$, $G^{'}_\a$, $G^{''}_\a$ and the modified amplifier $\amp{\a,\theta}(U)$ will reduce to $F$, $G$, $G^\dag$ and $\amp{}(U)$ in Appendix~\ref{appendix:amp}, respectively;
    \item [-] one can set $\a = \cos(\Delta) / \sin(m \Delta)$, $\theta = m\Delta$ to make Equation~\eqref{eqn:property mod amp} hold.
\end{enumerate}
\end{remark}

\subsection{Example: Qubit-unitary inverse} \label{appendix:qubit}
In this section, we provide an exemplary implementation of QURA in the qubit case $d = 2$. 
We start with $\decod(U)$.

\begin{figure}[H]
\centering
\[
\Qcircuit @C=0.8em @R=2.3em {
    & \qw &\qw & \ctrl{2} &\qw & \qw & \ctrl{2}  &    \qw & \qw \\
     & \qw  & \ctrl{1} & \qw  & \qw  & \ctrl{1}& \qw &\qw  & \qw \\
   & \qw  & \gate{Z} & \gate{X}  & \gate{\Uin} & \gate{Z} & \gate{X} & \qw& \qw}
\]
\caption{Circuit implementation of $\decod(U)$ given in Equation~\eqref{eqn:decoder}.}\label{fig:5 3 G}
\end{figure}

As for $\encod(U)$, since $U^* = Y U Y$ for every qubit untiary $U$, there does not need additional qubit to implement conjugate of $U$. $\ft_2$ is the Hadamard gate. Then $\encod(U)$ is constructed as

\begin{figure}[H]
\centering
\[
\Qcircuit @C=0.6em @R=1.5em {
    & \qw & \gate{H} & \qw & \ctrl{2} &\qw& \qw & \qw & \qw & \ctrl{2} &  \qw   &  \gate{H} & \qw \\
     & \qw & \gate{H} & \ctrl{1} & \qw  & \qw & \qw & \qw  & \ctrl{1} &\qw &\qw  &  \gate{H} & \qw \\
   & \qw & \qw & \gate{X} & \gate{Z} & \gate{Y} & \gate{\Uin} & \gate{Y} & \gate{X} & \gate{Z} & \qw &  \qw & \qw}
\]
\caption{Circuit implementation of $\encod(U)$ given in Equation~\eqref{eqn:encoder}.}\label{fig:5 3 QU}
\end{figure}

In the qubit case, $\Delta_d = \arcsin(1/2) = \pi / 6$ and hence $\pi /( 2 \Delta) = 3$ is an integer. Then there is no need to apply the modified amplitude amplifier at the end, and Equation~\eqref{eqn:approx} is reduced to $\amp{}^2 (\idop \ox \encod(U)) (\ket{000} \ox \ket{\psi}) = \ket{000} \ox U^\dag\ket{\psi}$.
The overall implementation is given as

\begin{figure}[H]
\[
\Qcircuit @C=1em @R=1em {
    & \lstick{\ket{0}} & \qw & \qw& \gate{X} & \ctrl{1} & \qw & \gate{X} & \qw & \ctrl{1} & \gate{X} & \qw & \qw & \qw_{\times 2} & \rstick{\ket{0}} \qw \\
    & \lstick{\ket{0}} & \multigate{2}{\encod(U)} & \qw & \ctrlo{-1} & \multigate{1}{F} & \qw & \multigate{2}{\decod(U)} & \qw & \multigate{1}{F^{\dagger}} & \ctrlo{-1} & \multigate{2}{\encod(U)} & \qw & \qw & \rstick{\ket{0}} \qw \\
    & \lstick{\ket{0}} & \ghost{\encod(U)} & \qw & \ctrlo{-1} & \ghost{F} & \qw & \ghost{\decod(U)} & \qw & \ghost{F^{\dagger}} & \ctrlo{-1} & \ghost{\encod(U)} & \qw & \qw & \rstick{\ket{0}} \qw \\
    & \lstick{\ket{\varphi}} & \ghost{\encod(U)} & \qw & \qw & \qw & \qw & \ghost{\decod(U)} & \qw & \qw & \qw & \ghost{\encod(U)} & \qw & \qw & \rstick{U^{-1}\ket{\varphi}} \qw \gategroup{1}{5}{4}{12}{1.1em}{--}
}
\]
\caption{ Quantum circuit for implementing qubit-unitary inverse within 5 calls of $\Uin \in \su(2)$. 
The dashed block refers to the amplitude amplifier $\cA$, wherein $F$ satisfies $F\ket{00} = \ket{00^\perp}$, and the rotation angle $t$ is also given in Lemma~\ref{lem:mod}. In the $d=2$ case, $\cA_{\a}$ would be executed at least $m=2$ times, to obtain $U^{-1}$.} \label{fig:5 3 deterministic}
\end{figure}

\section{Extensions}~\label{appendix:qura extension}

In this section, we discuss how QURA changes when the unitary $U$ to be reversed has some known structure. 
 
\subsection{Reversing controlled evolutions}~\label{appendix:control gate}
It is obvious to extend QURA to invert a control gate with less cost, regarding such control gate as a superposition of those controlled gates. For $U_0,\ U_2,\cdots U_{n-1}\in\su(d)$, the control gate, or the \emph{select gate} of $U_j$s is defined as
\begin{equation}
    \SEL_{U_\cdot}=\sum_{j=0}^{n-1} U_j \ox \ketbra{j}{j}_{n} \in \su(nd)
.\end{equation}
Then we have the following result.

\begin{corollary}
Let $m=\ceil{{\pi}/{2 \Delta}} -  1$, $\Delta = \arcsin{{1}/{d}}$, $\a=\cos\Delta/\sin(m\Delta)$, and $\SEL_{\amp{\a,\theta}(U)(U_\cdot)} = (\idop \ox \SEL_{\encod(U_\cdot)})\cdot(G_\a^{''} \ox \idop) \cdot (X \ox \SEL_{\decod(U_\cdot)})\cdot(G_\a^{'} \ox \idop)$. Then
\begin{equation}
    \SEL_{\amp{\a,m\Delta}(U_\cdot)}
    \SEL_{\amp{}(U)(U_\cdot)}^{m-1}
    (\idop\ox \SEL_{\encod(U_\cdot)}) \cdot 
    (\ket{0}_2 \ox \ket{00} \ox \idop_{nd}) 
    = \ket{0}_2 \ox \ket{00} \ox \SEL_{U_\cdot}^\dag
.\end{equation}
Moreover,
$\SEL_{\encod(U_\cdot)}$ consumes one call of $\SEL_{U_\cdot}^*$ and $\SEL_{\decod(U_\cdot)}$ consumes one call of $\SEL_{U_\cdot}$.
\end{corollary}

\subsection{Reversing multi-qudit evolutions}~\label{appendix:multi qubit}
In this section, we will extend Lemma~\ref{lem:oracle_d} into multi-qubit cases. Based on the new decomposition, those control-$X^j$ gates, control-$Z^j$ gates, and Fourier transform gates could be implemented more efficiently for reversing multi-qudit unitary operations.. Especially for multi-qubit cases, we could use control-$X$ gates, control-$Z$ gates, and $H$ gates instead of aforementioned gates.  Firstly, we assume such $d$-dimensional Hilbert space could be decomposed as a tensor product:
\begin{equation}
    \cH_d=\bigotimes_{s=1}^n \cH_{d_s}\text{ with } d=\prod_{s=1}^n {d_s}
.\end{equation}
In notation, we denote the decomposition of any computational basis $\ket{l}$ as
\begin{equation}
    \ket{l_1}_{d_1} \ox \ket{l_2}_{d_2} \ox \ldots \ox \ket{l_n}_{d_n} = \ket{l}
,\end{equation}
and three variant subcircuits as
\begin{align}
    \check{\encod}_{\ft}(U)& \coloneqq \sum_{j,k=0}^{d-1} \ketbra{j}{j} \ox \ketbra{k}{k} \ox \left(\left(\bigotimes_{s=1}^n Z_{d_s}^{j_s}X_{d_s}^{k_s}\right) \cdot U\cdot \left(\bigotimes_{s=1}^n Z_{d_s}^{-j_s}X_{d_s}^{k_s}\right)\right),\\
    \check{\encod}(U)& \coloneqq \left(\left(\bigotimes_{s=1}^n\ft_{d_s}^{\dag}\right)^{\ox2}\otimes \idop\right)\cdot \check{\encod}_{\ft}(U^*)\cdot\left(\left(\bigotimes_{s=1}^n\ft_{d_s}^{\dag}\right)^{\ox2}\otimes \idop\right),\\
    \check{\decod}(U)& \coloneqq \sum_{j,k=0}^{d-1}\ketbra{j}{j}\ox \ketbra{k}{k}\ox 
    \left(\left(\bigotimes_{s=1}^n X_{d_s}^{-j_s}Z_{d_s}^{-k_s}\right)\cdot U\cdot \left(\bigotimes_{s=1}^n X_{d_s}^{-j_s}Z_{d_s}^{k_s}\right)\right)
.\end{align}
Thus we have the following simplified dual relation similar to Corollary~\ref{cor:O1}, whose proof is routine and omitted here.

\begin{lemma}
\begin{align}
    &\check{\encod}(U)(\ket{00} \ox\idop)=\check{\decod}(U)^\dag (\ket{{+}{+}} \ox\idop),\\
    &\check{\decod}(U)(\ket{{+}{+}} \ox\idop)=\check{\encod}(U)^\dag (\ket{00} \ox\idop)
.\end{align}
\end{lemma}

By replacing $\encod(U)$ and $\decod(U)$ in Lemma~\ref{lem:mod} with $\check{\encod}(U)$ and $\check{\decod}(U)$, respectively, we obtain another quantum circuit to implement $\amp{\a,\theta}(U)(U)$, denoted as $\check{\cA}_{\a,\theta}(U)$. Finally, we have
\begin{equation}
    \check{\cA}_{\a,m\Delta}(U)\cdot \check{\cA}(U)^{m-1}\cdot (\idop\ox \encod(U))\cdot(\ket{0}_2 \ox\ket{00} \ox\idop)=\ket{0}_2 \ox\ket{00} \ox U^\dag,
\end{equation}
which is another implementation for $U^\dag$ similar as Theorem~\ref{thm:unitary inv cir exist}.

\subsection{Reversing evolutions with less complexity}~\label{appendix:less slots}

In this section, we provide a substitute to the amplitude amplifier $\widetilde{\cA}_{\a,m\Delta}(U)$ satisfying Equation~\eqref{eqn:approx} with less calls of $U$. For simplicity, we denote a set of unitary $\{V_{j,k}\}^{d^2}_{j, k=0}$ controlled by basis of two $d$-dimensional qudits $\{\ket{j, k}\}^{d^2}_{j, k=0}$ as
\begin{equation}
    \bigoplus_{j, k} V_{jk} \coloneqq \sum_{j,k=0}^{d-1} \ketbra{j, k}{j, k} \ox V_{jk}.
\end{equation}
Additionally, a unitary on the two $d$-dimensional qudits basis $\{\ket{j, k}\}^{d^2}_{j, k=0}$ is introduced as
\begin{equation}
    W\coloneqq\frac{1}{d} \sum_{j,k,l,r=0}^{d-1} \omega^{-(l-k)(r-j)} \ketbra{j, k}{l, r} \ox \idop.
\end{equation}
Denote $Q\coloneqq \left(\bigoplus_{j, k} Z^{-j} X^{-k} \right)\cdot  W \cdot\left(\bigoplus_{j, k} X^{2j} \right)$.
It is proved that

\begin{lemma}\label{lem:E-1}
For any $U=\sum_{\a,\beta=0}^{d-1}p_{\a\beta}Z^\a X^\beta\in\su(d)$,
\begin{equation}
    Q\cdot\decod(U)\cdot(\ket{{+}{+}} \ox\idop) = \sum_{\a,\beta=0}^{d-1}p_{\a\beta} \ket{\a, \beta}\ox\idop
.\end{equation}
\end{lemma}
\begin{proof}
Before the following proof, we point out that the decomposition of $U$ on Weyl operator basis is actually equivalent to Equation~\eqref{eqn:U_decom}. We first obtain
\begin{align}
    \decod(U)\cdot(\ket{{+}{+}} \ox\idop)
    =&\frac1d\sum_{j,k=0}^{d-1} \ket{j, k} \ox X^{-j}Z^{-k}UX^{-j}Z^k\\
    =&\frac1d\sum_{j,k,\a,\beta=0}^{d-1}p_{\a\beta} \ket{j, k} \ox X^{-j}Z^{-k}Z^\a X^\beta X^{-j}Z^k\\
    =&\frac1d\sum_{j,k,\a,\beta=0}^{d-1}\omega^{(\a-k)(\beta-j)}p_{\a\beta} \ket{j, k} \ox X^{\beta-2j}Z^{\a}.
\end{align}
Then the unitary operation $\sum_{j,k=0}^{d-1}\ketbra{j, k}{j, k}\ox X^{2j}$ on it leads to
\begin{align}
    &\left(\sum_{j,k=0}^{d-1}\ketbra{j, k}{j, k}\ox X^{2j}\right)\cdot\decod(U)\cdot(\ket{{+}{+}} \ox\idop)\\
    =&\left(\sum_{j,k=0}^{d-1}\ketbra{j, k}{j, k}\ox X^{2j}\right)\cdot\frac1d\sum_{j,k,\a,\beta=0}^{d-1}\omega^{(\a-k)(\beta-j)}p_{\a\beta} \ket{j, k} \ox X^{\beta-2j}Z^{\a}\\
    =&\frac1d\sum_{j,k,\a,\beta=0}^{d-1}\omega^{(\a-k)(\beta-j)}p_{\a\beta} \ket{j, k} \ox X^{\beta}Z^{\a}.
\end{align}
The designed unitary map $W$ eliminates phase terms,
\begin{align}
    &W \cdot\frac1d\sum_{j,k,\a,\beta=0}^{d-1}\omega^{(\a-k)(\beta-j)}p_{\a\beta} \ket{j, k} \ox X^{\beta}Z^{\a}\\
    =&\left(\frac{1}{d}\sum_{j,k,l,r=0}^{d-1}\omega^{-(j-r)(k-l)} \ketbra{l, r}{j, k} \otimes\idop\right)
    \cdot
    \frac1d\sum_{j,k,\a,\beta=0}^{d-1}\omega^{(\a-k)(\beta-j)}p_{\a\beta} \ket{j, k} \ox X^{\beta}Z^{\a}\\
    =&\frac{1}{d^2}\sum_{j,k,l,r,\a,\beta=0}^{d-1}\omega^{-(j-r)(k-l)+(\a-k)(\beta-j)}
    p_{\a\beta}\ket{l, r}\ox X^{\beta}Z^{\a}\\
    =&\frac{1}{d^2}\sum_{j,k,l,r,\a,\beta=0}^{d-1}\omega^{(l-\a)j+(r-\beta)k+\a\beta-lr}p_{\a\beta}
    \ket{l, r}\ox X^{\beta}Z^{\a}\\
    =&\sum_{l,r,\a,\beta=0}^{d-1}\delta_{l=\a}\delta_{r=\beta}\omega^{\a\beta-lr}p_{\a\beta}
    \ket{l, r}\ox X^{\beta}Z^{\a}\\
    =&\sum_{\a,\beta=0}^{d-1}p_{\a\beta}
    \ket{\a, \beta}\ox X^{\beta}Z^{\a}.
\end{align}
Finally, it is proved that
\begin{align}
    &\left(\sum_{jk}\ketbra{j, k}{j, k}\ox Z^{-j}X^{-k}\right)\cdot
    \sum_{\a,\beta=0}^{d-1}p_{\a\beta}
    \ket{\a, \beta}\ox X^{\beta}Z^{\a}\\
    =&
    \sum_{\a,\beta=0}^{d-1}p_{\a\beta}
    \ket{\a, \beta}\ox Z^{-\a}X^{-\beta}X^{\beta}Z^{\a}\\
    =&
    \sum_{\a,\beta=0}^{d-1}p_{\a\beta}
    \ket{\a, \beta}\ox\idop.
\end{align}
\end{proof}

\begin{remark}
The fixed gate $W$ defined in Lemma~\ref{lem:E-1} could also be written as 
\begin{equation}
    W= \frac{1}{\sqrt{d}}\sum_{j,k=0}^{d-1}\ket{j, k}(X^k\ft^\dag X^{-j})^{T_1},
\end{equation}
where $T_1$ is the partial transpose map satisfying $(\ketbra{l}{r})^{T_1}=\bra{l, r}$.
\end{remark}

Hence, Theorem~\ref{thm:unitary inv cir exist} is adjusted to:

\begin{theorem}\label{thm:unitary inv cir exist_mod}
There exists a quantum circuit to implement $U^{\dag}$ by $m$ calls of $U$ and $m$ calls of $U^*$ for arbitrary $U \in \su(d)$, where $m=\ceil{{\pi}/{2 \Delta}} -  1$, $\Delta = \arcsin{{1}/{d}}$, e.g.
\begin{equation}
    \widetilde{\cA}_{\a,m\Delta}(U)\cdot \amp{}(U)(U)^{m-1}\cdot (\idop\ox \encod(U))\cdot(\ket{0}_2 \ox\ket{00} \ox\idop) = \ket{0}_2 \ox \sum_{\a,\beta=0}^{d-1} p_{\a\beta}
    \ket{\a, \beta}\ox U^\dag,
\end{equation}
where 
\begin{align}
    \widetilde{\cA}_{\a,m\Delta}(U)& \coloneqq\left(\idop_2\ox Q\right)\cdot(G_\a^{''}\ox\idop)\cdot(X\ox \decod(U))\cdot(G_\a^{'}\ox\idop),\\
    G_\a^{'}&\coloneqq (\ketbra{1}{1}_2 \ox F_\a + \ketbra{0}{0}_2 \ox\idop)\cdot( (X_2-\idop) \ox \ketbra{00}{00}+\idop_2 \ox \idop),\\
    G_{\a}^{''}&\coloneqq (({Ry(t)}-\idop)\ox\ketbra{00}{00}+\idop)\cdot(\ketbra{1}{1}_2 \ox F^\dag+\ketbra{0}{0}\ox\idop),\\
    t&=-2\arctan\frac{1}{\tan\left(m\Delta\right)\sqrt{1-\a^2}},\\
    \a&=\cos\Delta/\sin(m\Delta).
\end{align}
\end{theorem}
\begin{proof}
Theorem~\ref{thm:unitary inv cir exist} has claimed 
\begin{equation}
    \amp{\a,m\Delta}(U)\cdot \amp{}(U)(U)^{m-1}\cdot (\idop_2\ox \encod(U))\cdot(\ket{0}_2\ox\ket{00} \ox\idop)=\ket{0}_2 \ox\ket{00} \ox U^\dag.
\end{equation}
Considering 
\begin{equation}
    \widetilde{\cA}_{\a,m\Delta}(U)=\left(\idop_2\ox Q\cdot\encod(U)^\dag\right)\cdot\cA_{\a,m\Delta}(U),
\end{equation} 
    when we multiply both sides of the above equation by the matrix $\idop_2\ox Q\cdot\encod(U)^\dag$, and we obtain
\begin{align}
    &\widetilde{\cA}_{\a,m\Delta}(U)\cdot \amp{}(U)(U)^{m-1}\cdot (\idop\ox \encod(U))\cdot(\ket{0}_2 \ox\ket{00} \ox\idop)\\
    =&\left(\idop_2\ox Q\cdot\encod(U)^\dag\right)\cdot\left(\ket{0}_2 \ox\ket{00} \ox U^\dag\right)\\
    =&\ket{0}_2 \ox Q\cdot\encod(U)^\dag\cdot\left(\ket{00} \ox \idop \right)\cdot U^\dag\\
    \xlongequal{\eqref{eqn:s_dual_2}}
    &\ket{0}_2 \ox Q\cdot\decod(U)\cdot\left(\ket{{+}{+}} \ox \idop \right)\cdot U^\dag\\
    \xlongequal{Lemma~\ref{lem:E-1}}
    &\ket{0}_2 \ox\left(\sum_{\a,\beta=0}^{d-1}p_{\a\beta}
    \ket{\a, \beta}\ox\idop\right)\cdot U^\dag\\
    =&\ket{0}_2 \ox\sum_{\a,\beta=0}^{d-1}p_{\a\beta}
    \ket{\a, \beta}\ox U^\dag.
\end{align}
\end{proof}

\begin{remark}
With the replacement of the final $\cA_{\a,m\Delta}(U)$ into $\widetilde{\cA}_{\a,m\Delta}(U)$, we reduce one call of $\encod(U)$ but the output two qudits no longer return $\ket{00} $. An exemplary qubit case of the modified circuit can be found in~\cite{mo2025parameterized}. 
\end{remark}

\end{document}